\documentstyle[preprint,aps]{revtex}   

\renewcommand{\today}{11th February  1999
\\
Revised: 16 March 1999}

\newcommand{\nc}{\newcommand}
\nc{\be}{\begin{equation}}
\nc{\ee}{\end{equation}}
\nc{\bea}{\begin{eqnarray}}
\nc{\eea}{\end{eqnarray}}
\nc{\beas}{\begin{eqnarray*}}
\nc{\eeas}{\end{eqnarray*}}
\nc{\noi}{\noindent}
\nc{\sD}{\not \! \! D}
\nc{\s}[1]{\not \! #1}
\nc{\non}{\nonumber}
\nc{\bb}{\bibitem}
\nc{\lf}{\left}
\nc{\ri}{\right}
\nc{\mb}[1]{\makebox[#1]{}}
\nc{\pa}{\partial}
\nc{\sA}{\not \! \! A}
\nc{\newsec}[1]{\section{#1}\mb{0.5cm}}
\nc{\h}{\frac{1}{2}}
\nc{\ra}{\rightarrow}
\nc{\la}{\leftarrow}
\nc{\ep}{$e^+e^-\ra\pi^+\pi^-\;$}
\nc{\emuon}{$e^+e^-\ra\mu^+\mu^-\;$}
\nc{\epp}{$e^+e^-\ra\pi^+\pi^0\pi^-\;$}
\nc{\elec}{$e^+e^-\ra\gamma^*\ra e^+e^-\;$}
\def\mathunderaccent#1{\let\theaccent#1\mathpalette\putaccentunder}
\def\putaccentunder#1#2{\oalign{$#1#2$\crcr\hidewidth
\vbox to.2ex{\hbox{$#1\theaccent{}$}\vss}\hidewidth}}

\nc{\ti}{\mathunderaccent\tilde}
\nc{\M}{{\cal M}}
\nc{\rw}{$\rho\!-\!\omega\;$}
\def\hhht{\rule[ 0.mm]{0.mm}{6.mm}}

\def\hhhb{\rule[-3.mm]{0.mm}{9.mm}}
\def\hhhc{\rule[-3.mm]{0.mm}{3.mm}}
\def\hhhu{\rule[-3.mm]{0.mm}{12.mm}}

\begin{document}
\tightenlines    
\draft           
\preprint{\vbox{Phys. Rev. {\bf D59}, 114027 (1999) \hfill UK/TP 99--01  \\
                                       \null \hfill LPNHE 99--01 \\
                                       \null \hfill SLAC--PUB--8048 \\
                                       \null\hfill hep-ph/9902326}}
\title{Radiative Decays, Nonet Symmetry and SU(3) Breaking}
\author{M.~Benayoun$^{a}$, L. DelBuono$^{a}$,
        S. Eidelman$^{a,b}$, V. N. Ivanchenko$^{a,b}$, 
H.B. O'Connell$^{c,d}$\thanks{Work supported by 
         Department of Energy contract DE-AC03-76SF00515.}}
\address{$^{a}$LPNHE des Universit\'es Paris VI et VII--IN2P3, Paris,
         France\\
         $^{b}$Budker Institute of Nuclear Physics, Novosibirsk 630090, 
         Russia \\
         $^{c}$Department of Physics and Astronomy,
         University of Kentucky,
         Lexington, KY 40506, USA\\
         $^{d}$Stanford Linear Accelerator Center, Stanford University,
         Stanford CA 94309, USA}
\date{\today}
\maketitle
\begin{abstract}
We re-examine the problem of 
simultaneously
describing in a consistent way all 
radiative and leptonic decays of light mesons ($V \ra P \gamma$,
$P \ra V \gamma$, $P \ra \gamma \gamma$, $V \ra e^+ e^-$).
For this purpose, we rely on the Hidden Local Symmetry model
in both its anomalous and non--anomalous sectors. We show that
the SU(3) symmetry breaking scheme  proposed by Bando, Kugo and Yamawaki,
supplemented with nonet symmetry breaking in the pseudoscalar
sector, allows one to reach a nice agreement with all data,
except for the $K^{*\pm}$ radiative decay. 
An extension of
this breaking pattern allows one to
account 
for this particular decay mode too. 
Considered together, the whole set of radiative decays 
provides a pseudoscalar mixing angle  $\theta_P \simeq -11^{\circ}$
and a value for $\theta_V$ which is $\simeq 3^{\circ}$ from 
that of ideal mixing.
We also show that it is impossible, in a practical sense, to disentangle 
the effects of nonet symmetry breaking
and those of glue inside the $\eta'$, using only light meson decays.
 
\end{abstract}

\newpage

\hspace{8.cm} When in doubt, tell the truth \ldots \\

\hspace{9.cm} Mark Twain

\pagenumbering{arabic}
\section{Introduction.}

Defining an accurate framework in which the study of all
radiative decays of light flavor mesons can be performed 
successfully is a long standing question. There are
a few kinds of different models which have been proposed so far.
The most popular modelling is in terms of magnetic moments
of quarks \cite{magn1,magn2}.
It includes, to some extent, SU(3) breaking effects by having
a magnetic moment for the $s$ quark, slightly different to that
of the $d$ quark. It also depends on 
overlap integrals which are
hard to estimate theoretically, and are thus
arbitrarily 
chosen  equal \cite{dolinsky}. This, at least,  allows one to have
a reasonable number of free parameters.
Another traditional approach is to use SU(3) relations 
among coupling constants
\cite{odonnel}. However, the assumption of exact SU(3) symmetry still
yields reasonable descriptions of radiative decays \cite{ben2},
though the success is never complete.

Recently, several models including SU(3) symmetry breaking effects
have been proposed \cite{BGP,hash,BES}, motivated 
in part by effective Lagrangian
approaches to the interactions of vector mesons \cite{HLS,FKTUY}, 
sometimes with
additional SU(2) symmetry breaking effects\cite{hash}. 
Including 
SU(3) symmetry breaking as per Bando, Kugo and Yamawaki 
(hereafter referred to as BKY) 
\cite{BKY,heath}, these models give a special role to all decay constants
($f_K$, $f_{\eta}$, $f_{\eta'}$) in the breaking procedure; some additional
corrections are also allowed.  More recently,
a new kind of model has been proposed  \cite{ball}, where additional symmetry
breaking effects are introduced by means of the (measured) leptonic
decay constants of vector mesons.

The study of radiative decays of light flavor mesons is also 
connected with the long standing problem of $\eta/\eta'$ mixing 
\cite{ben2,BES,ben1,chan,DHL,GILMAN} 
and to its possible association with a glue content\cite{ball,veneziano}.
The two--photon decay of the $\eta(\eta')$ is also connected
with the problem of anomalies \cite{veneziano,ABJ,WZW1,WZW2}. Recent
developments seem to advocate a more complicated $\eta/\eta'$ mixing scheme
\cite{leutw,leutwb}, which has received support from some phenomenological 
analyses
\cite{kroll1,kroll2}. 
 Another approach in the same vein
has been proposed
quite recently by Escribano and Fr\`ere \cite{frere}, where
the pseudoscalar mixing angle $\theta_P$ is taken to
be mass dependent. Practically, this
leads to two different mixing angles 
$\theta_{\eta}=\theta_P(m_{\eta}^2)$
and $\theta_{\eta'}=\theta_P(m_{\eta'}^2)$ in order to express the physical
states $|\eta\rangle$ and $|\eta'\rangle$ in terms of the octet and singlet 
pseudoscalar fields.
As a consequence of this assumption, however, the physical 
states $|\eta>$ and $|\eta'>$
are no longer orthogonal to each other. Similar discussions can be found for
$\pi^0-\eta$ mixing \cite{pieta} and 
\rw mixing \cite{rw}, where the latter is a necessary result of
current conservation \cite{node}.

Among the models sketched so far, only the O'Donnell model \cite{odonnel} 
introduces, quite naturally, the effect of nonet symmetry breaking,
simply by its being SU(3) (but {\it not} U(3)) symmetric in both the
pseudoscalar and vector sectors.
 As phenomenology tends to indicate the relevance of
deviations from nonet symmetry, it looks interesting to keep
this possibility, in addition to conventional SU(3) breaking effects.

The purpose of the present paper is to propose a new model based on the
successful approach of hidden local symmetry \cite{HLS}, and particularly
its anomalous sector \cite{FKTUY}. As we know that effects of SU(3)
symmetry breaking are clearly observed in the data \cite{ben2}, 
they have surely to be introduced.
We do it following the BKY mechanism \cite{BKY}. This approach reveals 
an interesting pattern for the $s-q$ breaking 
($q$ or $q'$ stand for $u$ or $d$ quarks).
Indeed, decays involving a  ($q \overline{q'}$) pseudoscalar meson 
have unchanged
coupling constants (they depend directly on $f_{\pi}$), while
decays involving a  ($s \overline{q}$) or a ($\overline{s}q$)
pseudoscalar meson are renormalized by $ f_{\pi}/f_K$,
and the ($s \overline{s}$)
part (only) of neutral meson couplings is corrected by $( f_{\pi}/f_K)^2$, 
without any change for the additional  ($q \overline{q}$) parts. 

This is automatically performed by a field renormalization \cite{BKY,heath}, 
which must
be propagated  \cite{heath} into the interaction 
(non--anomalous and anomalous) Lagrangians.
This breaking scheme has the virtue of being both
simple and successfully predicting the magnitude
and the functional form of the breaking effects. In order to probe various
kinds of effects, we supplement the BKY breaking by another one 
introduced by Bramon, Grau and Pancheri \cite{BGP} 
(hereafter referred to as BGP)
which acts directly on the anomalous Lagrangian of Ref.~\cite{FKTUY}.
Finally, we introduce there the possibility of breaking the nonet 
symmetry in the 
pseudoscalar sector.

The net result of this breaking pattern is threefold. 
Firstly, to explicitly construct
a Lagrangian model which gives the SU(3) model of Ref.~\cite{odonnel}
for $VP\gamma$ transitions.  
Secondly to perform consistently the BKY and BGP breakings,  
and thirdly to
test against data each of the elementary breaking schemes (BKY, BGP
and nonet symmetry)
separately. Finally, we will also examine the effects of 
a glue component coupled  to the $\eta$ and $\eta'$ mesons.

On another hand, using standard Feynman rules, this model also
allows  us to reconstruct the amplitudes for the anomalous decays 
$\eta/\eta' \rightarrow \gamma \gamma$.
It so happens that it provides the traditional mixing scheme with only one 
mixing angle $\theta_P$ and
with, additionally, expressions for the singlet and octet decay 
constants $f_1$ and $f_8$ 
in terms of $f_{\pi}$, $f_K$ and the nonet symmetry breaking parameter (named $x$). 
We thus have an access to testing whether radiative decays of 
type  $VP\gamma$ give support to this traditional scheme \cite{chan,DHL,GILMAN,WZW1,WZW2}, 
or if there is evidence for
its failure, in connection with the recent modelling involving 
two mixing angles \cite{leutw,leutwb,kroll1,kroll2,frere}.

The paper is organized as follows. In Section \ref{data}, we sketch
the present status of experimental data. In Section \ref{exactsymm}
we summarize the standard exact SU(3) model  of O'Donnell for radiative
decays and examine how the interplay of an additional singlet can be performed.
Section \ref{HLSmodel} is devoted to a brief reminder of the main tools~:
the HLS model, its anomalous sector and the SU(3) symmetry breaking schemes.
It is here (Subsection \ref{anomalous}) that we exhibit the Lagrangian model we use 
in order to describe the  radiative and leptonic decays of light flavor mesons. 
We discuss here, and in Section \ref{radlept},
the way nonet symmetry breaking is implemented and the connection between
the symmetry breaking parameters and the usual singlet and octet decay constants 
$f_1$ and $f_8$. In Section \ref{fitres}, we discuss the results of the fit
using the proposed model. The analysis of nonet symmetry breaking versus glue
component in the $\eta$ and $\eta'$ mesons is discussed in Section \ref{nsglue}.
Problems connected with the Primakoff measurement  of $\eta$ decay width are
shortly discussed in Section \ref{prima}. Predictions for branching fractions
are presented in Sections \ref{prediction} and \ref{prediction2}.
Section \ref{kstar} deals with a special treatment for including the $K^{*\pm}$
radiative decay mode inside the set of partial widths submitted to fit.
Section \ref{conclud} is devoted to conclusions. For ease of reading, most of the 
numerical results are gathered in  Tables and most formulae are given in the Appendix.

\section{The Data}
\label{data}

Recently there have been a few improvements of the data we analyze. Most
of them  have already been listed in Review of Particle Properties \cite{PDG98}. 
This concerns some decay modes like $\rho^0 \rightarrow \eta \gamma$ and 
$\rho^0 \rightarrow \pi^0 \gamma$
which were  previously  somewhat overestimated \cite{nsk}.  
Results have also been improved for the mode
$\omega \rightarrow \eta \gamma$ and new results have increased the quality
of the information on $\phi$ radiative decay to $\eta$ \cite{phieta1,phieta2}. 
Moreover, the decay mode $\phi \rightarrow \eta' \gamma$ has been measured
by the two detectors mounted on the VEPP--2M collider. The result of the
CMD2 collaboration \cite{phietp1} is the present reference \cite{PDG98} and
corresponds to a branching ratio of 
$(1.2^{+0.7}_{-0.5})\times10^{-4}$. More recently,
the SND collaboration has reported \cite{phietp2} 
a slightly smaller (but consistent within errors) branching ratio of
$(6.7^{+3.4}_{-2.9})\times10^{-5}$.

Therefore, we have at our disposal the full set of radiative decays of 
light mesons.
One might, however, expect potential difficulties with some particular data. 
The first is the disagreement between the Primakoff effect
measurement of $\Gamma(\eta \rightarrow \gamma \gamma)$, and the results
which have been obtained from six $e^+ e^-$ experiments (see  Ref.~\cite{PDG98}
for references).  One should note that 
the results provided by all $e^+ e^-$ experiments are statistically consistent
with each other (mean value~: $0.514 \pm 0.026$ keV)
whereas the Primakoff effect measurement ($0.324 \pm 0.046$ keV) 
is statistically
inconsistent with $e^+ e^-$ experiments by more than $3 \sigma$. Therefore,
it could be considered 
unreliable to mix the results from $e^+ e^-$ data with those from the
Primakoff effect. Hence, in the following, we
use the mean value of the  $e^+ e^-$ measurements  ($0.514 \pm 0.026$ keV), 
which  seems more reliable, due to the number of experiments. We shall,
however, somehow discuss the physical consequences of the  
Primakoff measurement ($0.324 \pm 0.046$ keV).

The second (potential) problem is the self--consistency of the measured 
values for the partial widths $K^{*\pm} \rightarrow K^{\pm} \gamma$ and
$K^{*0} \rightarrow K^{0} \gamma$ which
have been  recognized hard to reconcile with each other
by several authors. For instance the model of exact SU(3) symmetry \cite{odonnel}
provides \cite{ben2} a ratio 
$\Gamma(K^{*0} \rightarrow K^0 \gamma)/\Gamma(K^{*\pm} \rightarrow K^{\pm} \gamma)=4$,
as the (SU(3) broken symmetry) model of  Ref. \cite{ball}, while the reported data 
\cite{PDG98} rather give $2.2 \pm 0.3$. On the other hand, the model using quark magnetic
moments predicts \cite{volodia} a ratio of 1.65. In these last two cases,
the reported value for $\Gamma(K^{*0} \rightarrow K^0 \gamma)$ is well reproduced,
while the reported information \cite{PDG98} for $\Gamma(K^{*\pm} \rightarrow K^{\pm} \gamma)$
is twice larger compared to the prediction
of Ref. \cite{ball} or 50\% smaller compared to the prediction of Ref. \cite{volodia}.  
It is hard to determine the source of this discrepancy, which could
conceivably be  either due to systematic effects in the
measurements or simply reflect the need for a refined modelling.

\section{An Exact SU(3) Symmetry Framework}
\label{exactsymm}

\indent \indent The formalism which describes the decays 
$V \rightarrow P ~\gamma$
and $ P \rightarrow V \gamma$ within an exact  SU(3) symmetry framework 
has been given by P. O'Donnell in Ref.~\cite{odonnel}. 
The corresponding decay amplitudes can be quite generally{\footnote{ 
That is, independently of any specific Lagrangian.}} written as
\begin{equation}
T=g_{VP\gamma} \epsilon_{\mu \nu \rho \sigma}k^{\mu} q^{\nu} 
\varepsilon^{\rho}(V)\varepsilon^{\sigma}(\gamma) 
\label{model1} 
\end{equation}

\noindent
using obvious notations. The coupling strengths $g_{VP\gamma}$
between physical vector and pseudoscalar mesons in radiative decays
are expressed in terms of two angles ($\theta_V$ and $\theta_{P}$)
which describe the mixtures of singlet and octet components,
and of three coupling constants ($g_{V_8P_8\gamma}$, $g_{V_1P_8\gamma}$
and $g_{V_8P_1\gamma}$); indeed, assuming that the photon behaves like an SU(3)
octet cancels out the possible coupling $g_{V_1P_1\gamma}$.
We do not reproduce here the expressions for the  $g_{VP\gamma}$
in terms of the elementary couplings $g_{V_iP_j\gamma}$ and the mixing
angles~; they can be found in Ref.~\cite{odonnel} and 
in Appendix A7 of Ref.~\cite{ben2}, where a misprint has been corrected.
These formulae use mixing angles describing deviations from
ideal mixing, introduced long ago in Ref.~\cite{BG}, 
rather than the more customary mixing angles relative 
to octet and singlet components. The relevance of this angle definition 
has recently been rediscovered \cite{kroll1,kroll2} in connection with the
$\eta/\eta'$ mixing problem.

At this point, it should be noted that exact SU(3) symmetry is 
not in conflict with releasing the condition of nonet symmetry
(which corresponds to the stronger U(3) symmetry)
usually stated in effective Lagrangian models for both the vector and 
pseudoscalar meson sectors \cite{BES,HLS,FKTUY,HS,klingl}.
Moreover, the O'Donnell formulation also treats the mixing angles as
free parameters to be determined{\footnote{The relation of mixing angles 
to meson masses, and thus to some SU(3) breaking effects, is ignored,
allowing one to compare fit values with theoretical expectations.}}. 

Then, assuming the existence of only one singlet for each of the vector 
and pseudoscalar 
meson sectors, exact SU(3) symmetry gives a description which 
depends generally on five parameters. Reducing the number
of free parameters requires additional symmetries. In most
traditional approaches to vector meson physics, nonet symmetry
in the vector sector is assumed \cite{BES,HLS,FKTUY,HS,klingl}
once the vector meson field matrix has been written with ideal
$\omega$  and $\phi$ fields. A recent phenomenological study
of light meson radiative decays \cite{ben2} has concluded
that all existing data highly favor nonet symmetry in the  
vector meson sector. Practically, this means that one can limit oneself
to only two independent couplings~: $G=g_{V_8P_8\gamma}=g_{V_1P_8\gamma}$
and $G'_1=g_{V_8P_1\gamma}$~; if we define $x$ by $G'_1=xG$,
the result of the most likely fits to the data in Ref.~\cite{ben2}
(the so--called ``internal fit''  and ``model M1''),
tells that $x \simeq 0.9$, with a (statistical) error
making $x$ fully inconsistent with 1. This would imply
that nonet symmetry is significantly broken in the pseudoscalar
sector. Whether this value for $x$ is a fundamental property
of radiative decays, or an effective way to account for 
(ignored) SU(3) symmetry breaking effects, is still an open
question and partly motivates the present paper. 

The question of whether nonet symmetry in the pseudoscalar
sector is fulfilled is intimately related to the U(1)$_A$ anomaly and 
recent accounts of this problem can be found in Refs. \cite{grunberg,leutw,leutwb}. 
At large $N_c$, the nine basic pseudoscalar mesons should 
form a nonet of degenerate
Goldstone bosons, and then one could expect  that
the parameter $x$ defined above should tend to 1.
However, the precise value of $x$ should also be related to
the scale dependence of the singlet coupling constant 
\cite{veneziano,leutw,leutwb}. Therefore, from a phenomenological
point of view, it looks wise to let $x$ vary and examine
the consequences of assigning it specific values (including 1).
 
Moreover, recent theoretical developments tend to advocate
that the singlet sector of pseudoscalar mesons could well be
not saturated \cite{ball,veneziano,leutw,grunberg} by the 
standard singlet
$v_1=(u \overline{u}+d \overline{d}+s \overline{s})/\sqrt{3}$  only.
One (or more) of the glueballs predicted by QCD could play
a non--negligible role. Let us denote this additional state $v'_1=gg$,
exhibiting its possible connection with glue. Actually such a state
could be a compound of several SU(3) singlets, like several glueballs
with, even, a possible admixture of $c \overline{c}$, as recently
advocated in order to account for the unexpectedly high rate of
$B \rightarrow \eta' K$ observed at CLEO\cite{cleo} (see
Refs. \cite{ccbar,kroll1} and references quoted therein). 
The possibility that a part of the broad structure seen
in radiative $J/\psi$ decays and
presently named\cite{PDG98} $\eta(1440)$ could be a nearly pure 
glueball is still considered\cite{ochs} 
(see Ref.~\cite{amsler} and the minireview in Ref.~\cite{PDG98}).

Then, it is meaningful to allow for the mixing of 
$v_8=(u \overline{u}+d \overline{d}-2 s \overline{s})/\sqrt{6}$
with both singlet states already referred to as $v_1$ and $v'_1$.
This follows the proposal in Ref.~\cite{ball}. We
are not actually very dependent on an assumption about the precise content 
of $v'_1$, except that it is supposed to be orthogonal to $v_1$. 

It is suitable, for later use, to choose a 
parametrization of the mixing of $(v_8, v_1, v'_1)$ into physical pseudoscalar
meson states denoted $(\eta, \eta', \eta'')$.
Using the symbol $\eta''$ for the third partner
of the doublet $(\eta, \eta')$ simply means that we don't care to identify it ,
as  we do not presently plan to describe its coupling to vector mesons and photons
and, more generally, its physics. Any general  parametrization of an orthogonal
rotation matrix depends {\it a priori} on 3 angles. One could for instance choose 
to express it in terms 
of the usual Euler angles,
however, an appropriate parametrization of this transform is 
represented by the Cabibbo--Kobayashi--Maskawa matrix (with  the complex phase 
factor $\delta$ removed)
\begin{equation}
\left[
     \begin{array}{ll}
     \displaystyle \eta\\[0.5cm]
     \displaystyle \eta'\\[0.5cm]
     \displaystyle \eta''\\
     \end{array}
\right]
=
\left[
     \begin{array}{lll}
\displaystyle \cos{\theta}\cos{\beta} & -\displaystyle \sin{\theta}\cos{\beta} & \sin{\beta}\\[0.5cm]
\displaystyle \sin{\theta}\cos{\gamma}-\cos{\theta}\sin{\beta}\sin{\gamma} &
\displaystyle \cos{\theta}\cos{\gamma}+\sin{\theta}\sin{\beta}\sin{\gamma} &
\displaystyle \cos{\beta}\sin{\gamma}\\[0.5cm]
-\displaystyle \sin{\theta}\sin{\gamma}-\cos{\theta}\sin{\beta}\cos{\gamma} &
-\displaystyle \cos{\theta}\sin{\gamma}+\sin{\theta}\sin{\beta}\cos{\gamma} &
\displaystyle \cos{\beta}\cos{\gamma}\\
     \end{array}
\right]
\left[
     \begin{array}{ll}
     v_8\\[0.5cm]
     v_1\\[0.5cm]
     v'_1\\
     \end{array}
\right]
\label{model2}
\end{equation}

Indeed, the vanishing of  $\beta$ and $\gamma$ gives smoothly the usual mixing
pattern of the $(\eta,\eta')$ doublet (with $\theta\equiv \theta_P$)
and the decoupling of the additional singlet
(which will be frequently -- and abusively --named glue). 
Setting $\beta=0$
cancels out glue inside $\eta$ only, while $\gamma=0$ removes any glue
inside the $\eta'$ only. 

\section{The HLS Model and SU(3) Symmetry Breaking}
\label{HLSmodel}  

In order to break SU(3) symmetry relations among coupling
constants, it is convenient to refer to a well defined framework, in which
clear meaning can be ascribed to each parameter, particularly to 
parameters which define the breaking procedure itself. In this way,
fit parameters and values can be physically interpreted, which is
a missing information noticed \cite{ball} for the parameters{\footnote
{These parameters have  already been renamed with the corresponding 
capital letters in order to avoid ambiguities with other meanings
for the same symbols in use in the field of effective Lagrangian models.}} 
$g$, $ g_1$ and $g'_1$ of Ref. \cite{ben2}.

\subsection{The Unbroken HLS Model}

\indent \indent We will partly work within the framework of the hidden local 
symmetry
model (referred to in the following as HLS). A full account of this can be 
found
in Refs. \cite{HLS,BKY} and, for what concerns the anomalous sector, also 
in Ref. \cite{FKTUY}. Brief accounts can be found in 
Refs.~\cite{BGP,hash,heath,HS}
and will not be repeated here. However, in order to fix -- and modify -- 
notations, 
a few points have to be recalled. The HLS Lagrangian can be written
${\cal L}_{\rm HLS}={\cal L}_{A}+a {\cal L}_V$ where
\bea\non
{\cal L}_{A}&=&-\frac{f_{P}^2}{4}{\rm Tr}\left[
D_\mu\xi_L\xi_L^{\dagger}-D_\mu\xi_
R\xi_R^{\dagger}\right]^2
\\
{\cal L}_V&=&-\frac{f_{P}^2}{4}{\rm Tr}\left[
D_\mu\xi_L\xi_L^{\dagger}+D_\mu\xi_R\xi_R^{\dagger}\right]^2
\label{HLS1}
\eea
and $a$ is a parameter which is not fixed by the theory.
Setting $a=2$ allows one to recover the usual expression for
vector meson dominance (VMD) \cite{HLS} (for a review of
VMD see, for example Ref.~\cite{review}). However, some
experimental evidence \cite{pipi,cmd2} indicates that $a$
is slightly (but significantly) greater{\footnote{Ref.
\cite{jain} reached a similar conclusion when analyzing
resonance parameters in a model close to HLS.}} than 2. 
For this reason, we prefer to keep  $a$ free. 
After removal of a scalar field matrix, we have
\begin{equation}
\xi_L^{\dagger}=\xi_R=\xi=e^{iP(x)/f_{P}}
\label{HLS2}
\end{equation}
where $P$ is the pseudoscalar field matrix and $f_{P}$
is usually identified with the pion decay constant 
$f_{\pi}=92.42$ MeV\cite{PDG98},
at least for phenomenological purposes.
The HLS Lagrangian above is gauged for both electromagnetism
and the hidden local symmetry through the covariant derivative
\be
D_\mu\xi_{L,R}=\pa_\mu\xi_{L,R} -igV_\mu\xi_{L,R}+ie\xi_{L,R}A_\mu Q
\label{HLS3}
\ee
where $A_\mu$ is the electromagnetic field and $Q=$diag$(2/3,-1/3,-1/3)$
is the quark charge matrix. $V$ is the vector meson field matrix
\be
V=\frac{1}{\sqrt{2}}
  \left( \begin{array}{ccc}
   (\rho^0+\omega^I)/\sqrt{2}  & \rho^+             &  K^{*+} \\
            \rho^-           & (-\rho^0+\omega^I)/\sqrt{2}    &  K^{*0} \\
            K^{*-}           & \overline{K}^{*0}  &  -\phi^I   \\
         \end{array}\label{vector}
  \right).
\label{HLS4}
\ee

Note the superscript $I$ for the $\omega$ and $\phi$ fields,
which reminds that these fields correspond to ideal mixing~;
note also the sign in front of $\phi^I$ which defines
$\phi^I\equiv-s\overline{s}$. There are some reason to think
that there is some admixture of non--strange quarks inside the observed
$\phi$ meson (for instance, in order to account for the large
branching fraction  $\phi \rightarrow \pi^+ \pi^- \pi^0$). 
This is quite traditionally treated by stating that the observed
doublet ($\omega,\phi$)  (denoted without superscripts)  is obtained 
by rotating the doublet ($\omega^I,\phi^I$) in the following way
(without any additional change of sign, if we define the ideal
$\phi$ as we did)
\begin{equation}
\left(
   \begin{array}{ll}
     \omega \\[0.5cm]
     \phi \\
   \end{array}
\right)
\left(
   \begin{array}{llll}
        ~~\cos{\delta_V} & \sin{\delta_V}   \\[0.5cm]
	-\sin{\delta_V} & \cos{\delta_V}  \\
    \end{array}
\right)
\left(
   \begin{array}{ll}
     \omega^I \\[0.5cm]
     \phi^I   \\
   \end{array}
\right)
\label{HLS5}
\end{equation}
Thus ideal mixing corresponds to $\delta_V=0$~; correspondingly,
the mixing angle of the $\omega, ~\phi$ system
with respect to their octet and singlet components is
\begin{equation}
\theta_V=\theta_0 +\delta_V~~~,~~~ \tan{\theta_0}=1/\sqrt{2}.
\label{HLS6}
\end{equation}

In terms of the singlet and octet fields, 
the pseudoscalar field matrix can be written
\be
P=\frac{1}{\sqrt{2}}
  \left( \begin{array}{ccc}
            \frac{1}{\sqrt{2}}\pi^0+\frac{1}{\sqrt{6}}\pi_8+
            \frac{1}{\sqrt{3}}\eta_0&\pi^+ &  K^+ \\
            \pi^-  & -\frac{1}{\sqrt{2}}\pi^0+\frac{1}{\sqrt{6}}\pi_8
            +\frac{1}{\sqrt{3}}\eta_0  &  K^0 \\
            K^-             &  \overline{K}^0  &
             -\sqrt{\frac{2}{3}}\pi_8 +\frac{1}{\sqrt{3}}\eta_0 \\
         \end{array} 
  \right),
\label{HLS7}  
\ee
where an explicit use of nonet symmetry fixes the relative
weights of the isoscalar terms. In a way analogous to the vector meson
case, the connection of  $\eta_0$ and $\pi_8$ with the 
observable $\eta$ and $\eta'$ is defined by a rotation ({\it i.e.}
defined by a single angle{\footnote{
 Ref. \cite{frere} prefers changing this standard definition
by introducing 
two mixing angles $\theta_{\eta}$ and $\theta_{\eta'}$ instead of $\theta_P$~;
we shall briefly comment on this point later on.
}})
\begin{equation}
\left(
   \begin{array}{l}
     \eta \\[0.5cm]
     \eta'\\
   \end{array}
\right)
\left(
   \begin{array}{ll}
        \cos{\theta_{P}} & -\sin{\theta_{P}} \\[0.5cm]
	\sin{\theta_{P}} & ~~\cos{\theta_{P}}  \\
    \end{array}
\right)
\left(
   \begin{array}{l}
      \pi_8   \\[0.5cm]
     \eta_0  \\
   \end{array}
\right)
\label{HLS8}
\end{equation}

We define also the relation between{\footnote{The states corresponding to 
$\omega^I$ and $\phi^I$ for the pseudoscalar isoscalar mesons are named
respectively $\eta_q$ and $\eta_s$ in Refs.~\cite{kroll1,kroll2}.}}
 $\eta$ and $\eta'$
and the non--strange ($\simeq \omega^I$) and strange  ($\simeq \phi^I$)
pseudoscalar field  combinations by relations analogous to  Eq.~(\ref{HLS5})
with a phase $\delta_P$ analogous to $\delta_V$ in Eq.~(\ref{HLS6}).

The unbroken (nonet symmetric) HLS Lagrangian is given explicitly in Ref. 
\cite{heath}. In order to stay consistent with the convention we have adopted 
above concerning the $\phi^I$ field, all coefficients of 
terms involving a single $\phi^I$ field in Ref. \cite{heath} must
change their sign.

\subsection{Nonet Symmetry Breaking}

As recalled in Section \ref{exactsymm}, there is 
no experimental or phenomenological reason to suspect any
failure of nonet ({\it i.e.} U(3)) symmetry in the vector meson sector.
However, we have also recalled that there are phenomenological and
theoretical reasons to suspect  that nonet symmetry in the 
pseudoscalar sector might not hold exactly. In order to test
this, it is wise to allow for its possible violation in
a way which permits a smooth connection from broken
to unbroken nonet symmetry. Having to introduce only one singlet combination 
(denoted $v_1$ in Section  \ref{exactsymm}) might also be 
questioned \cite{ball,veneziano} as commented above. 

The most straightforward way to include deviations from
nonet symmetry is to do it from the very beginning 
 in the HLS non--anomalous Lagrangian
by making the replacement $\eta_0 \longrightarrow x\eta_0$.
As  can be seen from Eq.~(A1) in Ref.~\cite{heath},
this does not influence the interaction terms (which do not 
contain any interaction with the singlet $\eta_0$), but only the
(omitted) kinetic energy term $\partial \pi_8 \partial \pi_8
+\partial \eta_0\partial \eta_0$ which then becomes
$\partial \pi_8 \partial \pi_8+x^2\partial \eta_0\partial \eta_0$.
This could imply a redefinition of the singlet decay constant as $f'=xf_P$
along the lines of Refs.~\cite{BGP,hajuj}. 
However, if we replace the $\pi^8$ and $\eta_0$ fields
by the physical $\eta$ and $\eta'$, this anyway produces
a term proportional to $\partial \eta \partial \eta'$
which goes to zero when $x \rightarrow 1$.
Moreover, as the mass associated with the singlet
is {\it a priori} different from that of the octet 
(see for instance Refs.~\cite{leutw2,leutw}),
the same rotation generates a term  of the form $m^2 \eta \eta'$
proportional to some mass squared (denoted $m^2$). When assuming $x=1$,
the diagonalization of this mass matrix gives a definition of the mixing
angle in terms of physical meson masses. If $x \ne 1$, the situation is
not completely clear.

Anyway, the problem we meet with $\eta$ and $\eta'$ 
 is tightly connected with the difficult $U(1)_A$ 
anomaly problem, and the present  work cannot pretend
to solve it. As our main concern is to build a Lagrangian model which 
allows one to go beyond elementary  SU(3) 
symmetry in describing
radiative decays,  we will simply assume that this part of 
the kinetic energy term can be suitably arranged, 
even when breaking nonet symmetry. 

{From} the point of view of phenomenology, we
could just as well introduce this specific breaking  
directly in the interaction Lagrangian
of relevance for our purpose (see below). When having  
to study possible effects of an additional singlet 
(named $v'_1$ above) in the radiative decays involving $\eta$ and $\eta'$, 
we will also assume that the corresponding
kinetic energy term can be safely introduced{\footnote{
At the computational level, it could be done by introducing a $g$ entry
in the meson field matrices which thus become $4\times 4$, filled
with zeros, except for the $gg$ location of the pseudoscalar matrix
which is filled with this additional term, 
and extending the $Q$ matrix by a fourth
diagonal entry with  zero charge.}}.

\subsection{SU(3) Breaking Mechanisms of the HLS Model}

Basically, the SU(3) breaking scheme we use has been introduced 
by Bando, Kugo and Yamawaki \cite{BKY} (referred to as BKY) 
and has given rise to a few variants \cite{BGP,heath}, 
as well as  to extension to SU(2) 
breaking \cite{hash}. We refer the reader to Refs.~\cite{BGP,BKY,heath} for 
detailed analyses of the properties of known variants of the BKY 
breaking scheme. 

In the following, we use basic consequences common to
the original BKY mechanism \cite{BKY}, its hermitized variant
and the so-called new scheme, both discussed in Ref.~\cite{heath}.
In these cases, SU(3) symmetry breaking defines
a renormalized  pseudoscalar field matrix $P'$ in terms
of the bare one $P$ given above by
\begin{equation} 
P'=X_A^{1/2} P X_A^{1/2},
\label{brk1}
\end{equation}
where the breaking matrix $X_A$ writes diag($1,~1,~1+c_A$)
and we have \cite{BKY,heath}
\begin{equation} 
\ell_A \equiv  1+c_A=\left(\frac{f_K}{f_{\pi}}\right)^2= 
1.495 \pm 0.030 ~~~,
\label{brk2}
\end{equation}

 It should be noted \cite{heath}, that the field renormalization  
(Eq.~(\ref{brk1})) is requested in order  to recover the charge normalization 
condition, $F_{K^+}(0)=1$, expected for the kaon form factor $F_{K^+}(s)$, even
in presence of SU(3) breaking. The numerical value just given is deduced from 
the experimental information quoted in Ref.~\cite{PDG98}.

Concerning vector mesons, apart from changing the couplings of $K^*$ and 
$\phi$ mesons to pseudoscalar pairs, SU(3) breaking modifies the vector
meson mass terms and
their coupling to the electromagnetic field in the following way
\begin{equation}
\begin{array}{ll} 
\cal{L}= \cdots +& \frac{1}{2} a f_{\pi}^2 g^2 \left[(\rho^0)^2+
(\omega^I)^2+\ell_V (\phi^I)^2 \right] \\[0.5cm]
~&\displaystyle - a e f_{\pi}^2 g \left[\rho^0 +\frac{1}{3}\omega^I+
\ell_V \frac{\sqrt{2}}{3}\phi^I\right].A 
+\cdots
\end{array}
\label{brk3}
\end{equation}
in terms of ideally mixed states. We have 
defined{\footnote{Notice the square in the definition
of $\ell_V$.}}  $\ell_V \equiv (1+c_V)^2$.
Using Eq.~(\ref{HLS5}), we can reexpress the HLS Lagrangian in terms of 
physical field combinations. The Lagrangian piece given in 
Eq.~(\ref{brk3}) thus becomes
\begin{equation}
\begin{array}{ll} 
\cal{L}=& 
a f_{\pi}^2 g^2 \left[ (\rho^0)^2 +(\cos^2{\delta_V}+\ell_V 
\sin^2{\delta_V})\omega^2+
(\sin^2{\delta_V}+\ell_V \cos^2{\delta_V}) \phi^2 \right]
+a f_{\pi}^2 g^2 \ell_V \omega . \phi \\[0.5cm]
~&\displaystyle - a e f_{\pi}^2 g \left[\rho^0 +
\frac{1}{3}(\cos{\delta_V}+\ell_V \sqrt{2}\sin{\delta_V})\omega
-\frac{1}{3}(\sin{\delta_V}-\ell_V \sqrt{2}\cos{\delta_V}) \phi \right].A 
\end{array}
\label{brk4}
\end{equation}
   
The  coefficients affecting the $\rho$, $\omega$ and $\phi$ fields in the 
last term, are
commonly denoted $-e f_{\rho \gamma}$, $-e f_{\omega \gamma}$ and 
$-e f_{\phi \gamma}$.
They are estimated from the vector meson decay widths to $e^+e^-$ by
\begin{equation}
\Gamma(V \rightarrow e^+e^-)= \frac{4 \pi \alpha_{em}^2}{3 m_V^3} 
|f_{V \gamma}|^2
\label{brk5}
\end{equation}
(for a recent discussion of the determination of the ``leptonic widths''
of vector mesons, see Ref.~\cite{GO}).
Following recent evidence \cite{pipi,cmd2} that $a$ could be somewhat
different from
2, it should be noted that $a$ influences  the description of these decays.

One should note the occurrence of a direct transition term $\omega . \phi$,
generated by the rotation $\delta_V$, which vanishes when the masses of the 
$\omega$ and $\phi$ mesons become equal (no SU(3) breaking). This term plays
an important role when computing some matrix elements, as will be 
illustrated below.

Some connection between the value of vector meson masses and the 
value of $c_V$
could be inferred \cite{BGP,BKY,heath}; however, given the possible 
definition dependence
of the measured values for vector meson masses\cite{klingl}, 
it cannot be considered 
secure {\it a priori}.
In most variants of the BKY breaking mechanism, $c_A$ and $c_V$ remain 
unrelated,
except for one of these variants \cite{heath} where the condition 
$c_A=a c_V$ looks desirable.   
The explicit expressions for the SU(3) broken HLS Lagrangians using 
several breaking schemes
can be found in Ref.~\cite{heath}.

\subsection{A Phenomenological Lagrangian for Radiative Decays}
\label{anomalous}

Following Ref.~\cite{FKTUY} (referred to hereafter as FKTUY),
the anomalous,  U(3) symmetric, Lagrangian describing
$PVV$ interactions and, 
together with Eqs.~(\ref{brk3}) and 
(\ref{brk4}), $P V \gamma$ and $P \gamma \gamma$ transitions is given by
\begin{equation}
{\cal L}=- \frac{3 g^2}{4 \pi^2 f_{\pi}} 
\epsilon^{\mu \nu \rho \sigma} {\rm Tr}[\pa_\mu V_\nu\pa_\rho V_\sigma P].
\label{wz1}
\end{equation}
where the coefficient is such that the $\pi^0 \rightarrow \gamma \gamma$
amplitude has the expression expected from the triangle anomaly 
\cite{ABJ,WZW1,WZW2}
($\alpha/\pi f_{\pi}$). The $V$ and $P$ matrices have been defined above
in terms of singlet and octet fields (pseudoscalars mesons) or of 
ideally mixed states (vector mesons).

One should mention that using the $V$ and $P$ matrices defined above,
allows one to recover the  $VP\gamma$ couplings of Ref.~\cite{odonnel} in the
case of nonet symmetry in vector and pseudoscalar sectors. The most
general form could be obtained by breaking quite generally  the U(3)
symmetry. This can be achieved by weighting the singlet part of
the matrix $V$ by a parameter $y$ and that of $P$ by another one, $x$.
Therefore, supplemented in this way, Eq. (\ref{wz1}) is an appropriate
Lagrangian representation for the amplitudes of Eq. (\ref{model1}),
in which breaking procedures can be clearly implemented. 

The field renormalization following the BKY SU(3) breaking should be propagated down to
the FKTUY Lagrangian  using Eq.~(\ref{brk1})
\begin{equation}
{\cal L}=- \frac{3 g^2}{4 \pi^2 f_{\pi}} 
\epsilon^{\mu \nu \rho \sigma} {\rm Tr}[\pa_\mu V_\nu\pa_\rho V_\sigma X_A^{-1/2}P'X_A^{-1/2}].
\label{wz2}
\end{equation}

Then, the VVP Lagrangian is changed in a definite way by the symmetry
breaking parameter $\ell_A$ defined above (see Eq.~(\ref{brk2})) and 
supposed to have a well understood numerical value 
(practically 1.5). We also allow
for slightly more freedom by introducing the possibility of another 
SU(3) breaking mechanism, specific to the anomalous 
Lagrangian, initially proposed by Bramon, 
Grau and Pancheri \cite{BGP} (referred to as BGP). 
This changes Eq.~(\ref{wz2}) to
\begin{equation}
{\cal L}=- \frac{3 g^2}{4 \pi^2 f_{\pi}} 
\epsilon^{\mu \nu \rho \sigma} {\rm Tr}[\pa_\mu V_\nu X_W\pa_\rho 
V_\sigma X_A^{-1/2}P'X_A^{-1/2}]
\label{wz3}
\end{equation}
where $X_W=$diag($1,~1,~1+c_W)$ is a new symmetry breaking matrix depending
on a new (free) parameter $\ell_W=1+c_W$. 
In Section \ref{radlept} we shall provide this parameter with
a physical meaning analogous to the relation 
between $\ell_A$ and $f_K/f_{\pi}$. 

The renormalized $P'$ matrix is given in Eq.~(\ref{HLS7}), except
that we allow for nonet symmetry breaking by making the replacement 
$\eta_0 \longrightarrow x \eta_0 $. 
Therefore, the Lagrangian which  we will first use
to describe radiative decays  depends {\it a priori}  
on the 2 (free) mixing angles and 3 breaking parameters: 
$\ell_A$ (expected to be of the order 1.5), $\ell_W$,
and $x$ (expected \cite{ben2} to be of the order 0.9, except if 
explicit SU(3) breaking is sufficient to fully restore nonet symmetry). 
The precise values of the HLS parameter $a$ and of $\ell_V$
only influence the description of the leptonic decays of neutral vector mesons.

The explicit form of this Lagrangian is given in the Appendix.
In principle, from this Lagrangian and the non--anomalous 
${\cal L}_V$ Lagrangian piece given
in Eq.~(\ref{brk4}), one is able to construct decay amplitudes 
for the processes 
$V \rightarrow P \gamma$, $P \rightarrow V \gamma$, $V \rightarrow e^+e^-$
and $P \rightarrow  \gamma \gamma$.

\section{Radiative and Leptonic Decays of Light Mesons}
\label{radlept}

\indent \indent Using the Lagrangian of Eq.~(\ref{wz3}) 
and the piece in Eq.~(\ref{brk4}),
we can get the coupling constants for the decays $V \rightarrow P \gamma$ and
$P  \rightarrow V \gamma$, given by Eqs.~(\ref{cc1}) and  (\ref{cc2}). 
They  are
related to the  partial decay widths through
\begin{equation}
\left\{
\begin{array}{ll}
\Gamma(V \rightarrow P \gamma)=&\displaystyle \frac{1}{96\pi}
\left[\frac{m_V^2-m_P^2}{m_V}\right]^3 \left|G_{VP\gamma} \right|^2\\[0.5cm]
\Gamma(P \rightarrow V \gamma)=&\displaystyle \frac{1}{32\pi}
\left[\frac{m_P^2-m_V^2}{m_P}\right]^3 \left|G_{VP\gamma} \right|^2
\end{array}
\right.
\label{wz4}
\end{equation}

Eq.~(\ref{cc3}) gives the connection between the main coupling parameter $G$
in Ref.~\cite{ben2} 
and more usual quantities like the pion decay constant $f_{\pi}$ 
and the universal
vector meson coupling $g$, introduced by the covariant 
derivative in Eq.~(\ref{HLS3}).

Some of these couplings are totally unaffected 
by any breaking process (such as all $G_{V \pi \gamma}$), while 
some are affected
only by broken nonet symmetry 
(like $G_{\rho^0 \eta \gamma}$ and $G_{\rho^0 \eta' \gamma}$).
In couplings involving both isoscalar vector and pseudoscalar mesons, 
the two SU(3)
breaking mechanisms accounted for above by $\ell_A$ (BKY) and  $\ell_W$ (BGP), 
come mixed together in a way which could prevent one 
from separately testing the relevance of 
the two  kinds of breakings ($Z=\ell_W/\ell_A$).  However, in the 
radiative decays of the $K^*$ mesons, 
one is able to extract information about these two parameters 
as they act differently for the two different 
charged modes.  This is  fortunate,
since, otherwise, there would be no way to disentangle what comes from
$\ell_A$ (and thus from $f_K$) from what comes from $\ell_W$.
In order to compare with recent modellings, we see
for instance that the relation 
$G_{\rho^0 \eta \gamma}/G_{\rho^0 \eta' \gamma}=\tan{\delta_P}$ 
\cite{kroll2} is modified
by nonet symmetry breaking.

One can check that the coupling constants
deduced from the Lagrangian in Eq.~(\ref{wz1}), supplemented
by the replacement $\eta_0 \longrightarrow x \eta_0$, are in perfect
agreement with the exact SU(3) model of O'Donnell \cite{odonnel}. 
Indeed, setting
$\ell_W=\ell_A=1$ in Eqs.~(\ref{cc1}) and (\ref{cc2}), allows one to recover
the expected expressions for the coupling constants \cite{odonnel,ben2},
when assuming nonet symmetry in the vector sector only. {From} this point of
view, the breaking scheme provided by the BKY and BGP mechanisms, happens
to be non--trivial, as can be seen from Eqs.~(\ref{cc1}) and (\ref{cc2}).
For instance,
it should be noted that decays involving $\eta$ and $\eta'$ mesons
are not simply rescaled in the breaking procedure, but that 
the breaking procedure treats differently strange and non--strange
contributions to the coupling constants.

Following standard rules, the same Lagrangian information allows
us to reconstruct definite expressions (see Eq.~(\ref{cc4})) 
for the two--photon couplings of the 
pseudoscalar mesons; these are related to partial widths by
\begin{equation}
\Gamma(X \rightarrow \gamma \gamma) =
\frac{M_X^3}{64 \pi} |G_{X\gamma \gamma}|^2~~~,~~X=\pi^0 , ~\eta, ~\eta '~~~.
\label{wz5}
\end{equation}
 
Let us first remark that the expression for $G_{\eta\gamma \gamma}$ 
compares well with the corresponding expression of Ref.~\cite{takizawa}
deduced from the Nambu--Jona--Lasinio model, showing that  breaking
parameters in this reference, originally expressed as 
functions of effective quark masses, 
also get an  expression{\footnote{This might indicate
that some relation exists connecting effective
masses of quarks and meson decay constants. 
 }} 
in terms of $f_{\pi}/f_K$. More interesting is that
we recover the traditional form for these amplitudes,
({\it i.e.} the one mixing angle expressions 
\cite{chan,DHL,GILMAN,WZW1,WZW2}).
Using these standard expressions, one indeed gets  
through identification
\begin{equation}
\begin{array}{ll}
\displaystyle \frac{f_{\pi}}{f_8}=\frac{5-2Z}{3}~~, & ~~~
\displaystyle \frac{f_{\pi}}{f_1}=\frac{5+Z}{6}x~~~,
\end{array}
\label{wz6}
\end{equation}
where $Z=\ell_W/\ell_A$.
This shows that, in the limit of SU(3) symmetry, we have $f_8=f_{\pi}$ and
$f_1=f_{\pi}/x$, and, that $f_1=f_8=f_{\pi}$ supposes  
that there is no symmetry breaking at all.
It is interesting to note that the FKTUY Lagrangian \cite{FKTUY}, 
broken as we propose, expresses
all decay constants in terms of $f_{\pi}$, $f_K$ , $x$ and 
the BGP breaking parameter $\ell_W$  
(if these last two quantities are  found to depart significantly from unity).
 It should anyway be noted that $f_1$ and $f_8$ get expressions
in terms of parameters which can be calibrated outside
the set of $P \rightarrow \gamma \gamma$ decays.  
Eq.~(\ref{wz6}) shows that the BGP breaking parameter
is connected to the relation between $f_8$, $f_K$ and $f_{\pi}$.


Concerning leptonic decay widths,
if one relies on the model described above and
in the Appendix, a combination of the $f_{V\gamma}$ is 
unaffected by SU(3) breaking and
can be written 
$f_{\omega \gamma} \cos{\delta_V} +f_{\phi \gamma} \sin{\delta_V} 
=f_{\rho \gamma}/3$, or alternatively
\begin{equation}
f_{\omega \gamma} G_{\omega \pi^0 \gamma}+f_{\phi \gamma} 
G_{\phi \pi^0 \gamma}=
f_{\rho \gamma}G_{\rho^0 \pi^0 \gamma}
\label{wz7}
\end{equation}
in terms  of measured quantities only. Correspondingly, the $X_V$ breaking
is fully concentrated in another independent combination  
$f_{\phi \gamma} \cos{\delta_V} -f_{\omega \gamma} \sin{\delta_V} 
=(\sqrt{2}f_{\rho \gamma}/3) \ell_V$. 
This relation can be rewritten
in terms of measured quantities only
\begin{equation}
\ell_V=\frac{1}{\sqrt{2}}\frac{
f_{\phi \gamma} G_{\omega \pi^0 \gamma}-f_{\omega \gamma} 
G_{\phi \pi^0 \gamma}}{
f_{\phi \gamma} G_{\phi \pi^0 \gamma}+f_{\omega \gamma} 
G_{\omega \pi^0 \gamma}}
\label{wz8}
\end{equation}

Therefore, one has, as an alternative 
to mass relations \cite{BKY}, a coupling
relation to define $\ell_V$. Additionally, the HLS parameter $a$
fulfills
\begin{equation}
a=\frac{e}{8 \pi^2 f_{\pi}^3} \frac{f_{\rho \gamma}}{G_{\rho^0 \pi \gamma}}
\label{wz9}
\end{equation}
and can be extracted in this way from data; however, it seems more accurate to perform 
a global fit of all coupling constants involving leptons or photons.

Before closing this section, one practical remark could be of interest.
When computing, for instance, the decay amplitude $\pi^0 \rightarrow \gamma \gamma$,
one is tempted to consider only the leading order terms which can be written
symbolically
\[ 
A(\pi^0 \longrightarrow [\rho^0 \ra\gamma] [\omega \ra\gamma])+
A(\pi^0 \longrightarrow [\rho^0 \ra\gamma] [\phi \ra\gamma])
\] 
and possibly add up the next order terms 
\[
A(\pi^0 \longrightarrow [\rho^0 \ra\gamma] [\omega\ra\phi\ra\gamma])+
A(\pi^0 \longrightarrow [\rho^0 \ra\gamma] [\phi \ra\omega \ra\gamma]), 
\]
by taking into account
the $\omega \leftrightarrow \phi$ direct transition term given in Eq.~(\ref{brk4}).
In doing this way, we reach the surprising conclusion that  the decay width 
$\pi^0 \rightarrow \gamma \gamma$ is affected by 
breaking the SU(3) symmetry.
Actually, this is an artefact which can be circumvented 
by summing up (formally) the full 
series of such $\omega \leftrightarrow \phi$ transitions and, in this case, 
we indeed get the third Eq.~(\ref{cc4}), exactly as if one uses the
$\omega^I$ and $\phi^I$ combinations
for virtual lines ({\it i.e.} for lines going from a decay vertex to
the transition to photons).
This remark applies in computing any decay amplitude
involving $\omega$ and $\phi$ legs hidden beneath each $\gamma$ external leg.

\section{Fitting Decays Modes with the Broken Model}
\label{fitres}
\indent \indent
In this section, we focus on the model for coupling constants
given by  Eqs.~(\ref{cc1}) to (\ref{cc5}), {\it i.e.} we make the assumption  
that there is no glue hidden inside the $\eta$ or $\eta'$ mesons. 
The quantities submitted to fit are the (measured)  coupling constants
which can be deduced from the partial widths in the Review of Particle
Properties \cite{PDG98}, using formulae recalled above.  

\subsection{Comments on Radiative $K^*$ Decay Measurements}

As commented above (see Section \ref{data}),
there are potential problems with both $K^*$ decay modes. 
Therefore, we have followed the strategy of performing fits
of all radiative decay modes except for these two. Then, the fit values
of the free parameters allow us to predict a value
for the partial widths $K^{*0} \rightarrow K^0 \gamma$
and $K^{*+} \rightarrow K^+ \gamma$, making it possible
to compare the $\chi^2$ distance of each these
predicted values to the corresponding measured values \cite{PDG98}.
In all fits, we have found that the prediction for  
$K^{*0} \rightarrow K^0 \gamma$ is in fairly good agreement
with the corresponding measurement, while the expected
value for $K^{*+} \rightarrow K^+ \gamma$ is always
at about $5\sigma$ from the accepted value \cite{PDG98},
casting some doubt on the reliability of this measurement. Therefore,
in the fits referred to hereafter, the process $K^{*+} \rightarrow K^+ \gamma$
has been removed.
We shall nevertheless reexamine this question in  detail in Section \ref{kstar}. 

\subsection{The HLS Parameter $a$ and the Breaking Parameter $\ell_V$
from Leptonic Decays}

Among the quantities we fit, most depend on only
$G$, $x$, $\ell_A$, $\ell_W$  and the mixing angles $\theta_V$ and $\theta_P$: these are
the radiative decays $V \rightarrow P \gamma$, 
$P \rightarrow V \gamma$, and the decays $P \rightarrow \gamma \gamma$,
and they represent the most important part of the broken symmetry model
we have built.

On another hand, the leptonic decay modes of the vector mesons are interesting 
because they open a new window for estimating the HLS parameter $a$ and the
breaking parameter $\ell_V=(1+c_V)^2$. We could, as well, 
have used the combination
in Eq.~(\ref{wz7}) which does depend neither on $\ell_V$ nor $a$.
We have, however, preferred keeping the three modes 
$\rho^0/\omega/\phi \rightarrow
e^+e^-$, in order to get new determinations of $\ell_V$ and $a$.

Without going into too much detail, it happens (as could been foreseen from
the very existence of Eqs.~(\ref{wz8}) and (\ref{wz9})), that values
for these parameters  do not 
depend in any way on assumptions about the other sectors, nor
even on the global fit quality level.

In this way, the values provided by both radiative and leptonic decays
for these parameters are
\begin{equation}
\begin{array}{ll}
a=2.50 \pm 0.03~~~, ~~~\ell_V=1.38 \pm 0.03~~~ (c_V=0.173 \pm 0.013)
\end{array}
\label{wz10}
\end{equation}
This value for $a$ (quite inconsistent with the standard VMD expectation $a=2$)
should be compared  with the value reported from fitting
$e^+ e^- \rightarrow \pi^+ \pi^-$
data of Ref.~\cite{barkov} in  Ref.~\cite{pipi} ($a=2.37 \pm 0.02$), and
with the fit of the preliminary data collected by the new 
CMD2 detector \cite{cmd2} $a=2.35 \pm 0.02$. The final analysis 
of  this last data set is now available \cite{logashenko} and 
provides  $a=2.38 \pm 0.02$.
 
Even if comparable, one observes a significant difference (about $3\sigma$) and
its origin is unclear. However, one should notice that the value of $a$ measured
from $e^+ e^- \rightarrow \pi^+ \pi^-$ is determined by the magnitude
of a non--resonant contribution to the scattering amplitude, while
the value in Eqs.~(\ref{wz10}) is determined by meson decays. This could be
a signal of an additional SU(2) breaking or of some systematic errors
hard to identify presently. 

The value for $c_V$ differs by a factor of two from what would be expected
if we use the mass formulae \cite{BKY}. However, it seems that the data
on radiative and leptonic decays favor the relation 
suggested 
in Ref. \cite{heath}, $c_A=a c_V$. It should be useful to check
whether or not this is a numerical accident, 
for instance, by a detailed study of
the annihilation processes $e^+ e^- \rightarrow K \overline{K}$.

\subsection{The SU(3) Breaking Parameters $\ell_A$ and $\ell_W$}

\indent \indent Two of the key parameters in the broken SU(3) model
we use in describing the radiative decays of light flavor mesons,
are the BKY parameter \cite{BKY,heath} $\ell_A$ expected to 
be equal to $(f_K/f_{\pi})^2$ and the BGP parameter  $\ell_W$.

As commented in the Appendix, the dependence upon $\ell_A$ and $\ell_W$
is always through their ratio (which is already referred
to as $Z$) except for the $K^{*0}$ decay 
mode{\footnote{We remind the reader, that $K^{*\pm}$ decay mode
is anyway inconsistent with the other data and has been removed 
from the fit set.}}.
Therefore, we expect large correlations between $\ell_A$ and $\ell_W$. 
{From} a
first fit with these parameters both free, we obtain  
\begin{equation}
\ell_W-1=c_W=(-0.17^{+0.59}_{-0.35} ) \times10^{-3}
\label{bgpbrk}
\end{equation}
with a nice fit probability ($\chi^2/$dof$=10.74/8$). Therefore
we have performed the fit setting  $\ell_W=1$ (or $c_W=0$, {\it i.e.} no BGP
breaking), and found 
unchanged fit quality ($\chi^2/$dof$=10.74/9$). Thus we
may conclude  that there is no need for the BGP breaking mechanism
at a level visible in the existing data{\footnote{See however Section
\ref{kstar}.}}. In this last fit, we also get
\begin{equation}
\ell_A=1.480^{+0.049}_{-0.047} 
\label{bkybrk}
\end{equation}
which is almost exactly the value expected from the known ratio $f_K/f_{\pi}$
(see Eq.~(\ref{brk2})). 
This gives, of course, a strong support to the breaking
mechanism proposed by Bando, Kugo and Yamawaki \cite{BKY,heath}.
 Indeed, the relation between $\ell_A$ and $f_K/f_{\pi}$, which is
mandatory within the BKY breaking scheme in order to fulfill $F_{K^+}(0)=1$,
even after SU(3) breaking,
is found here to hold numerically to quite a nice precision.
 It should be noted
that it comes together with either nonet symmetry breaking 
($x \ne 1$) {\it or} glue inside the system $\eta/\eta'$, 
as will be commented on below.

At this point, a comment is of relevance about the results reported in 
Ref.~\cite{BGP}.
Our result in 
Eq.~(\ref{bgpbrk}), means that taking into account all radiative decay
modes of light mesons, forces one to remove the data for $K^{*\pm}$ decay mode
as it is inconsistent with all the rest; 
in this case the BGP breaking practically cancels
out. By taking a part of the decay modes only, the relative inconsistency
of the data for both $K^*$'s is not obvious and explains the result of 
Ref.~\cite{BGP}.

One could consider the result in Eq.~(\ref{bkybrk}) 
as providing an interesting estimate of 
$f_K/f_{\pi}$, independent of measurements of $K$ and $\pi$ decays
\begin{equation}
\displaystyle \frac{f_K}{f_{\pi}}=1.217^{+0.021}_{-0.019}. 
\label{fkfpi}
\end{equation}

The results metioned above
suggest that one can reasonably  
{\em fix}  $\ell_A=1.50$ (at its physical value),
and remove the BGP breaking ($\ell_W=1$) which happens to be useless, as soon
as the BKY breaking is correctly set up. Then, the single free
breaking parameter which influences the coupling constants in radiative
decays, beside mixing angles, is the nonet symmetry breaking parameter $x$.

\subsection{The Nonet Symmetry Breaking Parameter $x$}

Therefore, the preliminary fits sketched above allow us to 
conclude
that the only actual free symmetry breaking parameter is 
$x$, once we do not consider
a coupling of the $\eta/\eta'$ doublet to glue. Stated otherwise, except for
the two mixing angles, we only have two free parameters to fit the data set,
as in the unbroken case \cite{ben2}.
One, named $G$, is connected with the vector meson universal coupling $g$, the other
is the nonet symmetry breaking  parameter $x$. 
The former  is clearly fundamental ($G$)
while it is uncertain whether or not the latter should be
considered fundamental.

We give in the first data column of Table \ref{one}, the fit results assuming
nonet symmetry ($x=1$), and leaving free all other 
model parameters. The best $\chi^2/$dof
we reach is 30/9, showing that the BKY and BGP breaking mechanisms alone are
unable to mimic a violation of nonet symmetry in the pseudoscalar sector. The
mixing angle is at a value frequently 
obtained in this case \cite{ben2,BES,ball,BES2,BES3};
however, the fit probability is low enough 
($5\times10^{-4}$), that the assumption
of full nonet symmetry can be considered sharply disfavored. 

In the second data column in Table \ref{one}, we display the fit results, 
assuming
{\it fixed} SU(3) breaking conditions ($\ell_A=(f_K/f_{\pi})^2$ and
$\ell_W=1$); in this case, we get 
instead a very nice fit probability ($44\%$).
This fit provides $x=0.92 \pm 0.02$, and indicates a significant departure
from nonet symmetry ($\simeq 4 \sigma$). More appealing is the mixing angle
of pseudoscalar mesons coming out from fit~: 
$\theta_P=-11.59^{\circ} \pm 0.76^{\circ}$, 
in perfect agreement with the linear mass formula, which predicts 
$-10.1^{\circ}$.

One should also notice that the vector mixing angle is found to be
$3.4^{\circ}$ smaller than its ideal value, in agreement
with the prediction of Ref.~\cite{DM} and previous fit results
\cite{ben2}, keeping in mind, however, that
this sign for  departure relative to ideal mixing depends
on the definition of the ideal $\phi$, 
and that we use $\phi^I=-s \overline{s}$.

The value for $G=0.704 \pm 0.002$ GeV$^{-1}$ 
is remarkably stable, independently of fit conditions, as can
be seen in Table \ref{one}.
We shall discuss later on (see Sections \ref{prediction} and \ref{prediction2}), 
other consequences
of the value found for $G$, which is also in perfect agreement
with the result obtained \cite{ben2} assuming no SU(3) breaking symmetry.
This is not surprising, as $G$ is  practically determined
by all $\rho$ decay modes and  the $\phi$ and $\omega$ decays to
$\pi^0 \gamma$, that is from a large number of processes not affected
by SU(3) breaking (see Eqs.~(\ref{cc1}) and (\ref{cc2})).

\subsection{The One Angle $\eta/\eta'$ Mixing Scheme from VMD}
\label{mixing}

As discussed above, and in the Appendix, the model we propose, which 
relies on the VMD approach of 
Refs.~\cite{HLS,FKTUY}, with fixed SU(3) breaking
\`a la BKY \cite{BKY,heath}, leads to  (one angle)
formulae for the $\eta/\eta' \rightarrow \gamma \gamma$ decay amplitudes, 
 which can be identified with the corresponding Current Algebra
standard expressions. This justifies the identification shown in 
Eq.~(\ref{wz6}) for the singlet and octet coupling constants.
One should note that nonet symmetry breaking does not conceptually
modify the formulae substantially.

In this case, we obtain together with  
$\theta_P=-11.59^{\circ} \pm 0.76^{\circ}$
\begin{equation}
\begin{array}{ll}
\displaystyle \frac{f_8}{f_{\pi}}=0.82 \pm 0.02~~~, & 
~~  \displaystyle \frac{f_1}{f_{\pi}}=1.15 \pm 0.02
\end{array}
\label{f1f8}
\end{equation}
using Eq.~(\ref{bkybrk}), and the fit result for $x$.

One should note that the value obtained for $f_8/f_{\pi}$
is not in agreement with the ChPT numerical expectation \cite{leutw2,GL85}.
This is a quite mechanical consequence of the low
value we get from our fit  for the mixing angle, as can be read off Fig. 1 in
Ref. \cite{ball}. 
However, this  does not prevent VMD from providing
quite a satisfactory description of all observables 
associated with light meson decays, including
a nice fit value of $\ell_A \simeq f_K/f_{\pi}$,  as seen
above and as will be illustrated below. 

One should also note that relatively low values
of $\theta_P$ have been advocated (or found) in analyzing similar data, 
for instance in Refs. \cite{ball,kroll1,BES3}. 
It can thus be remarked that
in the one mixing angle approach, it is only the addition of
$J/\psi$ decays which pushes $|\theta_P|$ to larger values.

\subsection{Radiative Decays versus $\gamma \gamma$ Decays}

The results reported in the previous subsections
mix $PV\gamma$ and $P\gamma \gamma$ couplings. In connection
with recent works \cite{leutw,leutwb,frere}, one could ask whether the good
description obtained 
using only one mixing angle, $\theta_P$, is not merely an artefact 
produced by merging these two kinds of couplings in a single fit procedure. 
In order to test this point, we have performed fits of the
$PV\gamma$ processes in isolation.

In a first fit, we removed the $K^{*0}$ mode, in order
to have a reference without any influence of $K^*$ decay modes.
Fixing, as usual{\footnote {In this case, we are actually only sensitive to the
{\em ratio} of breaking parameters $Z=\ell_W/\ell_A$.}} 
$\ell_A=1.5$ and $\ell_W=1$, we get a nice fit ($\chi^2/$dof$=8.6/7$),
finding that the observed value for $ K^{*0} \rightarrow K^0 \gamma$
was only $0.57 \sigma$ from the predicted value, while the data for
$ K^{*\pm} \rightarrow K^{\pm} \gamma$ was  $4.84 \sigma$ from
expectation. Therefore, it was indeed meaningful to include the
$ K^{*0}$ data in the fit, which is practically unchanged 
($\chi^2/$dof$=8.9/7$). 

In this case, we also found that the measured
value for $\eta \rightarrow \gamma \gamma$ was  $2.18 \sigma$
from the prediction of Eq.~(\ref{cc4}), while the data 
for $\eta' \rightarrow \gamma \gamma$ was $0.50 \sigma$ from
the corresponding prediction. In this last fit, the parameter values
we get are practically indistinguishable from the information in the
second data column of Table \ref{one}; the value for $\theta_P$ was, by
far, the most sensitive{\footnote{This simply reflects that 
Eqs.~(\ref{cc4}) only depend on the mixing angle.}} and moved
to --$10.41^{\circ} \pm 1.21^{\circ}$, that is only by  $\sim 0.5 \sigma$.



{From} this we can conclude that the two--photon decays for
$\eta$ and $\eta'$ are fully predicted from  $VP\gamma$
processes, which can be considered as a considerable success
of the FKTUY Lagrangian \cite{FKTUY}, supplemented by
the BKY breaking mechanism \cite{BKY,heath} set at a fixed
value, and of broken nonet symmetry. 

One might ask oneself whether our result about $f_8$ 
(which mechanically follows from simply the BKY breaking scheme)
is really in contradiction with the ChPT expectation $f_8\simeq 1.25 f_{\pi}$.
The answer is presently unclear because the definition we
use for the decay constants proceeds from the Wess--Zumino--Witten
Lagrangian \cite{WZW1,WZW2}, as reexpressed in Refs. \cite{chan,DHL,GILMAN},
and not on the coupling of pseudoscalar mesons to an axial field. 
This question will be examined in a forthcoming paper.


As a conclusion,  within the context of light meson decays,
we find no  failure of the VMD approach sketched above{\footnote{The 
problem with the $K^{*\pm}$
radiative decay can be solved {\it ad minima} as will be shown in Section
\ref{kstar}. }} and no need for a second angle \cite{leutw,leutwb,frere}
arises naturally from the data examined so far.  
However, one cannot 
exclude that nonet symmetry breaking is somehow equivalent to 
this second angle. This does not seem easy to prove from 
standard algebra.

\section{Nonet Symmetry Breaking versus Glue}
\label{nsglue}

\indent \indent Up to now, we have clearly illustrated that
the BKY breaking was a fundamental tool in order to describe
all data concerning radiative and two--photon decays of light
mesons. Correspondingly, we have shown that, at the level of
refinement allowed by the data, there was no need for the
additional BGP breaking.

Another central result of our fitting model concerns the
unavoidable need of about 10\% breaking of nonet symmetry 
in the pseudoscalar sector ($x \simeq 0.9$). Even if small,
this should be considered significant, as it is more than a $4 \sigma$ effect.
This could well be a property  that has to be 
accounted for at a fundamental level. However, nonet symmetry
breaking could well be an effective way to account (maybe partly at least) for
another physical effect, ignored in the model presented above.

We have already discussed the possible interplay of an additional
singlet component which could be present inside the $\eta/\eta'$ system.
This could be a gluonium component, 
a coupling to
$c \overline{c}$ or a mixture of both. It is not the purpose of the present 
paper to
try identifying this third component and the corresponding partner
of the $\eta$ and $\eta'$ mesons. Our concern is rather to see whether
a coupling to such an additional singlet could play some role in the
problem we examine.
 
The coupling to this additional state (hereafter named glue, somehow abusively)
has been presented in Section
\ref{exactsymm} and is summarized by Eq.~(\ref{model2}). The corresponding
coupling constants, as far as they are affected, are explicitly given
by Eqs.~(\ref{cc6}) to (\ref{cc8}).

A preliminary study of these relations, which include both SU(3)
breaking effects, nonet symmetry breaking and glue has been performed.
The conclusions are twofold~:

\begin{itemize}
\item The previous conclusions concerning the BKY and BGP breaking
mechanisms are unchanged. More precisely, the BKY breaking is found
determined by the value of $f_K/f_{\pi}$, while a possible BGP
breaking is found too small to be observed. Numerically, all
conclusions of the previous section remain fully valid.

\item Nonet symmetry breaking and glue are intimately connected
and reveal a correlation close to the 100\% level.

\end{itemize}

The second point does not mean that nonet symmetry breaking 
is physically equivalent to assuming coupling to glue, but
numerically it is indeed so. 
It also tells 
that, if we know what is the precise amount of glue, one can deduce
the level of nonet symmetry breaking (or conversely).

Table \ref{two} displays the main results of fitting the data
with a fixed level of symmetry breaking. We remind
that $\beta$ governs the level of glue inside the $\eta$ meson, while
$\gamma$ governs the level of glue inside the $\eta'$ meson.
It is clear from this table, that $x$ cannot be too small~;
for  $x \simeq 0.8$ and smaller values, glue is unable to account
for the two--photon decays{\footnote{At $x=0.7$ the predicted amplitudes
for $\eta/\eta' \rightarrow \gamma \gamma$ are both at $7 \sigma$ from
the measured values~; at $x=0.8$, the disagreement is still about 
$4 \sigma$ each.}} of $\eta$ and $\eta'$. Values larger than
$x \simeq 0.9$ look all statistically acceptable, except that
$\theta_P$ becomes less and less negative, and  even changes its sign! 
Moreover, the glue content needed in $\eta$ increases slowly from
zero (at $x=0.9$) to an arbitrary value. 

This clearly illustrates that above some level for $x$ (about 0.85),
one cannot distinguish the effects of glue from those of a genuine
nonet symmetry breaking effect. 

Two interesting features are however visible in Table \ref{two}~:

\begin{itemize}
\item  At $x \simeq 0.9$ both $\beta$ and $\gamma$ are quite consistent with
zero. This indicates that $x \simeq 0.9$ implies a decoupling of the glue
from $\eta$ and $\eta'$ mesons. This is the case developed in the preceding 
Section.

\item In the case of full nonet symmetry $x \simeq 1.0$, it is interesting 
to remark that $\beta$ is still consistent with zero, pointing to the fact that nonet symmetry
implies that $\eta$ couples only to the standard quarkonium states named $v_8$ and $v_1$
in Section \ref{exactsymm}. In this case however, the glue content of the $\eta'$ meson
becomes  significant.  
\end{itemize}

In view of these results, it looks justified to perform a fit,
fixing (as before) the SU(3) breaking effects to $\ell_A=1.5$, $\ell_W=1$,
$x=1$ and also $\beta=0$ (in order to lessen at most correlation effects).
In this case we have exactly the same number of parameters as
in the previous set of fits.  The corresponding fit results
are also displayed in Table \ref{one} and show a nice quality ($\chi^2/dof=10.5/10$),
equivalent to the no--glue case. 

The glue content this implies for the $\eta'$ meson can be presented in several ways.
Writing $\eta'=X v_8 +Y v_1+Z v_1'$ (with $X^2+Y^2+Z^2=1$), 
we have $\sqrt{X^2+Y^2}=0.89$ and $Z=0.46$. One can express the glue
fraction as $Z^2=\cos^2{\gamma}\simeq 0.20$ (at $x=1$).

As major conclusions of this section, one can first assert that a possible glue 
content inside the $\eta$ is not requested by the data. A significant
glue content inside the $\eta'$ is however subject to the actual level
of nonet symmetry breaking. 

We do not discuss any more values and meaning of $f_1$ and $f_8$.
Eq.~(\ref{cc8}), indeed shows that the meaning of these has to be revisited.
Moreover, the specific two--angle formulation of the $\eta /\eta' \rightarrow
\gamma \gamma$ decays introduced 
by the glue coupling ($\theta_P$ and $\gamma$), 
appears quite different from the one introduced in Ref. \cite{leutw,leutwb}.

\newcommand{\eee}{$e^+e^-\hspace{.2cm}$}
\section{Primakoff  versus \eee Information for $\eta$ Decay}
\label{prima}

All fit results presented so far, assumed the use
of the partial width $\eta \rightarrow \gamma \gamma$ as obtained
in $e^+e^-$ experiments. As mentioned above, this looks {\it a priori}
better grounded, as the number of $e^+e^-$ experiments is large~; 
this kind of  experiments looks also more straightforward
to interpret.

However, in case where the (single) measurement relying on the 
Primakoff  effect might have to be considered, it is not completely
useless to examine rapidly its consequences.

We have fit under the two assumptions of nonet symmetry breaking
(and no glue) and coupling to glue (with no nonet symmetry breaking).
All other parameters have been set at their usual values. The results are 
given in the last two data columns of Table \ref{one}.

The first remark is that the fits in both cases work as well as
when using the $e^+e^-$ information for $\eta$. A few undesirable features
are~:

\begin{itemize}
\item  The mixing angle $\theta_P$ is significantly smaller (in absolute magnitude)
with the Primakoff measurement than with the $e^+e^-$ data.

\item  Replacing nonet symmetry breaking by glue, leads to glue content in
$\eta$ as large as in the $\eta'$, if we estimate its fraction by the angles $\beta$ and 
$\gamma$).

\end{itemize}

The second item above might dismiss physically the reliability of
the Primakoff Effect measurement of $\eta \rightarrow \gamma \gamma$.

\section{Estimates for Branching Fractions from Fits}
\label{prediction}

\indent \indent The fits we have performed provide under various
symmetry breaking conditions, the parameter values and errors 
 given in Table \ref{one}. The cases  we will discuss here
correspond to the second and third data columns  in Table \ref{one},
which both give a very good fit quality. These are the cases with {\bf i/}
nonet symmetry breaking supplemented  by a fixed SU(3) breaking (BKY) 
and  {\bf ii/} a fixed SU(3) breaking (BKY) 
with glue inside the $\eta'$ replacing nonet symmetry breaking.
We now compare the branching fractions predicted by these two 
solutions to the accepted branching fractions as given in the
Review of Particle Properties \cite{PDG98}. They are computed
using the formulae for coupling constants given in the Appendix
and the relations defining the partial widths. The coupling constants
just referred to are computed from the basic parameters ($G$, $x$,
$\theta_P$, $\theta_V$, $\beta$, $\gamma$), by identifying these with gaussian distributions
having as mean values the central values in the fit and as standard
deviations, the corresponding ($1\sigma$) error.  

In Table \ref{three}, we first list the information for radiative decays only.
The first remark which comes to mind by comparing the two models
is that their predictions are close together.
The  relative disagreement with $\eta' \rightarrow \rho^0 \gamma$
is actually an artefact. Indeed, what has been fit is not 
the branching fraction given in Ref. \cite{PDG98}, but
the corresponding coupling constant which has been extracted
by the Crystal Barrel Collaboration in \cite{abele}. The reason
for this is that the (published) branching fraction for 
$\eta' \rightarrow \rho^0 \gamma$  is influenced by the box anomaly
\cite{ben2,chan,GILMAN,WZW1,WZW2} for the vertex
$\eta' \pi^+ \pi^- \gamma$ which is not accounted for in the VMD model
of \cite{FKTUY}~; actually this process contributes to the $\chi^2$
for only $\simeq 0.5$.

On the other hand, the PDG information reported for $\eta \ra \gamma \gamma$
branching fraction is the official one \cite{PDG98}, somehow influenced
by the Primakoff measurement. 

The single clear disagreement of model predictions with  data  
concerns the branching fraction for $K^{*\pm} \rightarrow K^{\pm} \gamma$,
that we find about half of the reported value
in the Review of Particle Properties \cite{PDG98}.
We postpone to Section \ref{kstar} the reexamination of this question. 

Otherwise, the largest disagreement is never
greater than about $1.5\sigma$. At such a (non--significant)
level, it is hard to distinguish whether differences
between predictions and data are due to SU(2)
breaking effects missing in the models, to systematic errors in the data
or to the (unavoidable) influence of the resonance models used 
to extract branching fractions from data.
For instance, changing the model for the $\rho$ lineshape
in the cross section for $e^+ e^- \rightarrow \pi^0 \gamma$
allows to reduce the branching ratio for $\rho^0 \rightarrow
\pi^0 \gamma$ from $(6.8 \pm 1.7)~10^{-4}$ to
$(6.1 \pm 1.5)~10^{-4}$ which compares better to the corresponding 
prediction ($5.2 ~10^{-4}$).

The new measurement for $\phi \rightarrow \eta' \gamma$
is also well accepted by the fit. However, the prediction
tends to indicate that the central value found by SND Collaboration
\cite{phietp2} is favored compared to that of the
CMD2 Collaboration \cite{phietp1}.

Additionally, Table \ref{four} shows that leptonic decays of
vector mesons are very well described. 

{From} all this, we can conclude that the model of symmetry
breaking, we have presented provides a consistent description
of the data. At their present level of accuracy, these
do not seem to require additional symmetry breaking effects.

An especially satisfactory conclusion is that  SU(3)
breaking effects are not left free in the fits and are
practically determined by the ratio $f_K/f_{\pi}$.
Some nonet symmetry breaking in the pseudoscalar sector
is, however, requested by the data. This is fully
or partly degenerated with a possible admixture of glue,
shown to (possibly) affect only the $\eta'$ meson.
If this has to be seriously considered, the question is to
identify the third partner to the ($\eta,\eta'$) doublet
which has been named $\eta''$. For this purpose, a
precise study of the decay properties of the $\eta(1440)$
meson could improve the hint. One has also to mention that
this glue component could be a $c \overline{c}$ admixture.

\section{Hadronic Decays of Vector Mesons}
\label{prediction2}

{From} the above fits of radiative decays, we get
stable estimates for the HLS parameters $a$ ($2.5 \pm 0.03$)
and $g$ ($5.65 \pm 0.02$) and also for the breaking parameter
$\ell_V$ ($1.38 \pm 0.03$). The deviation from ideal mixing
is  $\delta_V= -3.33 \pm 0.16$ degrees. In this Section,
we use the fir results obtained assuming nonet symmetry
breaking and SU(3) breaking with $\ell_A=1.5$ (second data
column in Table \ref{one}).

If one relies on the SU(3) broken HLS non--anomalous Lagrangian
\cite{heath}, using these parameter values, one can extract 
the coupling constants $g_{\rho \pi \pi}$ and also, for instance,
the coupling constants $g_{\phi K^+ K^-}$ and $g_{\phi K^0 \overline{K}^0}$
and compare to the corresponding data.

A value for $g$ can be extracted from fit to the $e^+e^- \ra \pi^+ \pi^-$
data \cite{pipi}, using $G_{\rho \pi \pi}=ag/2$ and one gets $g=5.18 \pm 0.02$.
This value compares poorly to the value we just get using
radiative decays of light mesons ($5.65 \pm 0.02$). This 
means  that, from radiative decays, one might expect a broader
width for the $\rho$ meson. However, the coupling constants for the
$\phi$ carry the same information as $\rho^0$. Indeed, it is easy to
get using Ref. \cite{heath}  

\begin{equation}
\displaystyle G_{\phi K \overline{K}}=-\frac{ag}{4\ell_A} 
\left[ \sin{\delta_V}-\sqrt{2} \ell_V \cos{\delta_V} \right]
\label{phicoupling}
\end{equation}
for both possible final states, $ K^+ K^-$ and $ K^0 \overline{K}^0$.
{From} standard formulae this coupling constant provides the 
branching fractions shown in Table \ref{four} which
compare quite well to the data. 

Therefore, in contrast with the hadronic width of the $\rho$,
the hadronic width of $\phi$ is in nice agreement with
the value found for $g$ when fitting radiative decays
($5.65 \pm 0.02$), with symmetry breaking parameters only influenced
by leptonic decays of vector mesons.

One can attempt the same comparison with $K^*$ hadronic decays
to $K \pi$ (as studied in Refs.~\cite{HS} and \cite{BGP2}). 
In this case, the coupling constants can be 
read off the Lagrangian given in Ref. \cite{heath}. Expliciting
the breaking parameter dependence, they are

\begin{equation}
\left\{
\begin{array}{lll}
\displaystyle g_{K^{*0}K^0\pi^0}&
\displaystyle =\frac{g_{K^{*0}K^+\pi^-}}{\sqrt{2}}&
\displaystyle =\frac{ag}{4}\sqrt{\frac{\ell_V}{\ell_A}}\\[0.5cm]
\displaystyle g_{K^{*\pm}K^{\pm}\pi^0}&
\displaystyle =\frac{g_{K^{*\pm}K^0\pi^{\pm}}}{\sqrt{2}}&
\displaystyle =\frac{ag}{4}\sqrt{\frac{\ell_V}{\ell_A}}
\end{array}
\right.
\end{equation}

Using the experimental data \cite{PDG98}, $\ell_A=1.5$,
and the fit values for the parameters $a$ and $\ell_V$, one
can extract $g=5.86\pm 0.10$ (neutral $K^*$) and  $g=5.98\pm 0.11$ 
(charged $K^*$) which
compare relatively well to our fit value ($5.65 \pm 0.02$),
even if it is not as good as for the $\phi$ meson hadronic decays.
 
This quite unexpected situation seems likely to be connected to the issue
of defining masses and widths for broad resonances like the $\rho$
and $K^*$ mesons. To be more precise, this addresses the question
of the connection between these physical parameters extracted
from data, obtained using Breit-Wigner resonance lineshapes, and the 
corresponding quantities occurring in a Lagrangian for broad resonances
\cite{BOW}.
This question will be studied in a forthcoming paper.
 
\section{The $K^{*\pm}$ Radiative Decay Problem}
\label{kstar}

\indent \indent The question of whether the radiative decay
$K^{*\pm} \rightarrow K^{\pm} \gamma$ is definitely
beyond the scope examined so far should be answered.

The first point which comes to mind is whether the disagreement
reported above (a factor of two between prediction and measurement) 
could be explained by breaking the SU(2) flavor symmetry.
The answer is seemingly no~; indeed, taking into account the quark content of the $K^*$'s, 
one could rather 
guess that a significant unaccounted for SU(2) breaking would
affect the quality of predictions for $K^{*0}$ rather
than for $K^{*\pm}$. However, the absolute partial width of the $K^{*0}$ is
well predicted by our modellings (flavor SU(3) and nonet symmetry breakings
and/or glue).
  
This possibility seeming unlikely, the question becomes~:
can the modelling developed in the Appendix be modified
in order to account for this mode within an extended SU(3) breaking 
framework? The reply is positive and is the following. 

Within the  spirit of the BKY mechanism, the (unbroken)
FKTUY Lagrangian given in Eq. (\ref{wz1}) can be broken
straightforwardly in three different ways. The first
mean is the pseudoscalar field renormalization
(see Eq. (\ref{wz2})), which leads to introduce the 
matrix $X_A$ and thus the breaking parameter
$\ell_A$ expected \cite{BKY} and found (see Eq. (\ref{bkybrk}))
equal to $(f_K/f_{\pi})^2$. It has been supplemented
with nonet symmetry breaking for reasons already presented
and with the success we saw.

The second mean is the BGP breaking \cite{BGP} of the anomalous 
FKTUY Lagrangian illustrated by Eq. (\ref{wz3}), which turns
out to introduce a breaking matrix $X_W$ and a new breaking
parameter $\ell_W$. Within the schemes presented up to
now, the fit value found for this parameter (practically $\ell_W=1$)
cancels out such a possibility, leaving
us with only{\footnote{Of course, the mixing angles are to be 
fit, as seen above.}}
 nonet symmetry breaking and the {\it fixed} original BKY
breaking scheme ($\ell_A \equiv (f_K/f_{\pi})^2$).

A third mean is however conceivable which has not yet been explored to our knowledge.
One should note that the BKY breaking mechanism \cite{BKY} implies a renormalization
(or redefinition) of the pseudoscalar field matrix expressed through $X_A$~; however,
the $X_V$ breaking does not end up with a renormalization of the vector field matrix, which
remains unchanged in the breaking procedure. In order to go on, let us
{\it postulate} that the vector meson field matrix has also to be SU(3) broken, 
and that this can be done by performing the change~:

\begin{equation}
V \longrightarrow X_T V X_T~~,~~~[~X_T=\rm{diag}(1,1,1+c_T)~]
\label{wzp1}
\end{equation}

\noindent in Eq. (\ref{wz3}), in complete analogy with the renormalization
of the $P$ matrix. One could as well assume that this breaking affects 
directly the FKTUY Lagrangian, however one cannot avoid remarking that
this looks indeed like a field renormalization, exactly as for $P$.
Doing this way, we have
a priori 3 parameters at our disposal in order to describe the full set of data of 
relevance~: $\ell_A, ~\ell_W$, 
and{\footnote{Notice the square in the definition for $~\ell_T$.
}} $~\ell_T=(1+c_T)^2$.

The Lagrangian corresponding to this case is given in Eq. (\ref{lagwzb}).
The coupling constants in Eqs. (\ref{cc1}) are unchanged in this new scheme
except for $K^*$'s which become~:

\begin{equation}
\left \{
\begin{array}{lll}
G_{K^{*0} K^0 \gamma}=&- & \displaystyle G \frac{\sqrt{K'}}{3} (1+\ell_W \ell_T )~ \\[0.3cm]
G_{K^{*\pm} K^{\pm} \gamma}=& &\displaystyle G \frac{\sqrt{K'}}{3}(2-\ell_W \ell_T)  
\end{array}
\right.
\label{wzp2}
\end{equation}

\noindent where $K'=\ell_T/\ell_A$. Eqs. (\ref{cc2}) and (\ref{cc3})
are also unchanged, provided one replaces there $Z$ by $Z'=Z \ell_T^2$.
Eqs. (\ref{cc5}) are, of course, unchanged{\footnote{
If $X_T$ actually follows from a renormalization of the vector
field matrix, it might  affect the expression for the leptonic decay widths 
considered. In this case, the values we get for $a$ and $\ell_V$
account effectively for this.}}.

Assuming no coupling to glue, we have performed the fit and found
a perfect fit quality ($\chi^2/dof=10.9/9$)~; in this case, the $K^{*\pm}$
is found to contribute to the full $\chi^2$ for $1.63 ~10^{-3}$,
while the  $K^{*0}$ contribution increases with respect to previous modelling
and becomes  0.68, {\it i.e.} unsignificantly. The fit parameters 
common to the previous (no--glue) model are found unchanged
(see second data column in Table \ref{one}), except for 
$\theta_P=-11.91^{\circ} \pm 1.10^{\circ}$, which thus has moved by
only $0.3\sigma$. This fit provides~:

\begin{equation}
\left \{
\begin{array}{lll}
 \ell_T=1.24 \pm 0.06& (c_T=~0.113 \pm 0.025) \\[0.3cm]
 \ell_W=0.66 \pm 0.06& (c_W=-0.340 \pm 0.058) \\[0.3cm]
\end{array}
\right.
\label{wzp3}
\end{equation}
 
\noindent with a large correlation coefficient $(\ell_T,\ell_W)=-0.935$.
This can easily be understood~: the pseudoscalar mixing angle is practically
fixed by the set of (single photon) radiative decays. Then, as shown indirectly
by Fig. 1 
in Ref. \cite{ball}, a mixing angle of $\theta_P \simeq -11^{\circ}$
forces the value of $Z'$ to be practically $1/\ell_A$~;
thus, $\eta/\eta'$ decays to two photons force  $\ell_T$ and $\ell_W$
to fulfill $\ell_W \ell_T^2 \simeq 1$. 

Therefore, it is indeed possible to accommodate all decays we examine.
However, there is an additional price to pay in order to include the $K^{*\pm}$ inside
the fit set of data~: we need to renormalize the vector field
matrix (introduction of $\ell_T$) and simultaneously break the FKTUY
Lagrangian \`a la BGP, in such way that these parameters 
fulfill an interesting condition

\begin{equation}
\ell_W \ell_T^2=1 
\label{wzp4}
\end{equation}

Stated otherwise~: accounting for this
measurement implies to include one more parameter. However,
this additional parameter plays in opposite directions
for both $K^*$ radiative decays as illustrated by
Eqs. (\ref{wzp2}). Assuming the validity of the
condition proposed by Eq. (\ref{wzp4}), implies
a highly non--trivial relation between the $K^{*\pm}$
and $K^{*0}$ radiative decay widths (or coupling constants),
which can hardly be accidental if it is fulfilled.
The relation just above between $\ell_W$ and $\ell_T$ 
is however a  numerical property and does not 
{\it a priori} mean that $X_W$ and $X_T$ are connected
from some basic principle.

Therefore we have redone the fit described just above,
requiring Eq. (\ref{wzp4}) among the corresponding fit parameters.
We got a good fit quality ($\chi^2/dof=11.07/10$) with exactly
the same parameter values as given in the second data column of Table \ref{one}
and additionally

\begin{equation}
\ell_T = 1.19 \pm 0.06~~~, ~~(c_T=0.109 \pm 0.024)
\label{wzp5}
\end{equation}

\noindent which turns to fix the BGP breaking parameter to $\ell_W=0.71 \pm 0.07$.

Thus, in order to account for $K^*$ radiative decays, two elementary breaking
mechanisms interplay with algebraically related strengths. Even if somehow non--trivial,
this procedure works without destroying the reconstruction quality of the $K^{*0}$ radiative 
decay, which was not obvious, knowing that we were looking for a factor of 2 for one only
of these two modes. The prediction of this last model for branching fractions are
listed in Tables \ref{three} and \ref{four} under the entry name ``$K^{*\pm}$ Breaking''.
All predictions are clearly in nice agreement with all accepted data \cite{PDG98}.

Nevertheless, this mechanism clearly complicates the full breaking picture
which is otherwise quite simple. One can hope that new measurements
for the $K^{*\pm}$ radiative decay will come soon from
the CLEO, BaBar and BELLE detectors through
the decay process $\tau^{\pm} \rightarrow K^{*\pm} \nu_{\tau}$.
This, anyway, would clarify the nature of the problem. 
If confirmed, this branching fraction  would rise the question of wether $X_T$ can be a 
renormalization of the vector 
field matrix~; if the relation between $X_T$ and $X_W$ (the BGP breaking)
is probably accidental, a connection between $X_T$ and $X_V$ (the second breaking
matrix in the BKY breaking mechanism) is not excluded.

We do not discuss at length, the correlation
between glue component and nonet symmetry, the conclusions reported above remain
fully valid. 

\section{Conclusion}
\label{conclud}

We have built a model aimed at describing
the radiative and leptonic decays of light flavor mesons,
including the two--photon decays of pseudoscalar mesons.
This model relies  heavily on the HLS model supplemented with
the BKY breaking mechanism in order to account for SU(3) symmetry breaking.
It has been shown that this is not enough to provide a satisfactory 
description of the available data, even by leaving free this breaking 
parameter and,  even, by allowing for an additional breaking scheme (BGP).

A nice agreement is however reached by allowing, additionally, either nonet
symmetry breaking in the pseudoscalar sector or having
an additional singlet which affects essentially the $\eta'$ meson.
This has been named glue, but could be as well any additional
kind of SU(3) singlet (a $c \overline{c}$ component for instance).

The picture that emerges from there is quite consistent
and tends to indicate that present data do not require
any breaking of the SU(2) symmetry at a visible level
in only radiative decays of light mesons.

We thus find that the radiative decays of the kind
$V \rightarrow P \gamma$ and $P  \rightarrow V \gamma$
allow one to predict quite precisely the decay widths
for $P \rightarrow \gamma \gamma$. We also
found, as expected from the BKY breaking mechanism,
that the relation between the breaking parameter and
$f_K/f_{\pi}$ is perfectly fit from data.

As a side result, we have shown that the HLS model
in its anomalous sector leads to the traditional one angle
mixing pattern for the ($\eta,\eta'$) system. In this case, 
the pseudoscalar mixing angle is $\theta_P=-11.59 \pm 0.76$ 
degrees in nice agreement with the value expected from linear 
mass formulae, but in poor agreement with ChPT expectations.  
This value for $\theta_P$ is, however, practically determined
by only the radiative one--photon decays of light flavor mesons
and therefore follows closely VMD expectations.

The single data which could require a special breaking procedure
is  the $K^{*\pm}$ radiative decay, if expected measurements confirm
the present accepted data. The existing measurements
can, however, be accommodated at the expense of
complicating somehow the SU(3) breaking scheme in our VMD model.
New measurements of this mode can be expected from 
$B$ factories in a near future~; this should 
tell definitely if such complications are really needed. 

Finally, we have shown that effects due to nonet symmetry
breaking and glue (and/or $c \overline{c}$ admixture)
cannot be practically disentangled, and then, in order to guarantee
presence or absence of glue inside light mesons, one has first
to ascertain the level of nonet symmetry breaking.

\vspace{1.0cm}
\begin{center}
{\bf Acknowledgements}
\end{center}
We thank J.-M. Fr\`ere (ULB, Brussels) for  
reading the manuscript and providing useful comments.
One of us (MB) acknowledges useful discussions 
with B. Moussalam (IPN, Orsay).
HOC was supported by the
US Department of Energy under contracts DE--FG02--96ER40989 (UK) and
DE--AC03--76SF00515 (SLAC). SE was supported by
the Division des Affaires Internationales of IN2P3 and would like to
thank
the LPNHE Laboratory for its hospitality; VNI was supported by
the Direction des Affaires Internationales of CNRS.
Both SE and VNI are grateful to
Eliane Perret (IN2P3) and
Marcel Banner (LPNHE) for their help and support.

\begin{table}[htb]

\begin{tabular}{|| c  | c  | c  | c | c |c ||}
\hline
\hline
\hhhb
      &\multicolumn{3}{c |}{$\eta  \rightarrow \gamma \gamma$} 
      & \multicolumn{2}{c |}{$\eta  \rightarrow \gamma \gamma$} \hhht \\
      &\multicolumn{3}{c |}{$(e^+e^- \rightarrow \eta e^+e^-$)} 
      & \multicolumn{2}{c |}{(Primakoff Effect)}\hhhb \\
\hline
\hline
\hhhc Breaking & SU(3) & Nonet Symm.&  Glue     &  Glue  &Nonet Symm. \\  
\hhhc Conditions & only      & $+$ SU(3)  &  $+$ SU(3)&  $+$ SU(3)&  $+$ SU(3)\\
\hline
\hline
$G$ [GeV]$^{-1}$\hhhb&
$0.703 \pm 0.002$ & $0.704 \pm 0.002$&
$0.704 \pm 0.002$ & $0.704 \pm 0.002$&$0.704 \pm 0.002$\\
\hline
$x$ & {\bf 1.~~}&
$0.917 \pm0.017$ & {\bf 1.~~} &
 {\bf 1.~~} &$0.854 \pm0.025$ \hhhu \\
\hline
\hline
$\beta$ [deg.]\hhhu&
{\bf 0.~~} & {\bf 0.~~}&
{\bf 0.~~}  & $21.34^{+3.43}_{-4.06}$ &{\bf 0.~~} \\
\hline
$\gamma$ [deg.]\hhhu&
{\bf 0.~~} & {\bf 0.~~} &
$26.38^{+2.51}_{-2.79} $ & $21.28^{+2.30}_{-2.36}$ &{\bf 0.~~} \\
\hline
\hline
$\ell_A$ \hhhu&
$1.41^{+0.60}_{-0.41}$ & {\bf 1.50}&
 {\bf 1.50}& {\bf 1.50}& $1.624 \pm 0.059$\\
\hline
\hline
$\ell_W$ \hhhu&
$[1-0.24~10^{-3}]^{+0.42}_{-0.30}$ & {\bf 1.~~} &
{\bf 1.~~} & {\bf 1.~~}& {\bf 1.~~} \\
\hline
$\ell_V$ \hhhu&
$1.376 \pm 0.031$ &$1.376 \pm 0.031$ &
$1.376 \pm 0.031$ &$1.376 \pm 0.031$ & $1.374 \pm 0.031$\\
\hline
$a$ [HLS] \hhhu&
$2.503\pm 0.033$ & $2.502 \pm 0.034$&
$2.503 \pm 0.034$ & $2.503 \pm 0.034$ & $2.506 \pm 0.034$\\
\hline
\hline
$\theta_V$[deg.]\hhhu & ~$31.92 \pm 0.17$ & ~$31.92 \pm 0.17$ & ~$31.92 \pm 0.17$ 
& ~$31.90 \pm 0.17$ & ~$31.88 \pm 0.17$\\
\hline
$\theta_P$[deg.]\hhhu & --$13.94 \pm 0.94$ & --$11.59 \pm 0.76$ & --$10.67 \pm 0.79$ 
&  --$~6.04 \pm 1.36$ &  --$~5.53 \pm 2.08$\\
\hline
\hline
$\chi^2/dof$ \hhhu & 31.9/9 & 10.9/10  & 11.1/10 & 6.3/9& 10.1/9\\
\hline
\hline
\end{tabular}
 

\caption{
\label{one}
Fit results under various strategies.
Parameter values written boldface means
that they are not allowed to vary. Nonet
Symmetry violation and glue in this table are exclusive of each other.
$K^{*+} \rightarrow K^+ \gamma$ is outside all fits;
its distance to the value expected from fit is always
about $5\sigma$.
}
 
 

\newpage 
 
\begin{tabular}{|| c  | c  | c |c | c |c ||}
\hline
\hline
\hhhc Breaking  &   & &  & &\\  
\hhhc Conditions & $x=0.7$    & $x=0.9$ & $x=1$ &$x= 1.1$ & $x= 1.3$\\
\hline
\hline
$\beta$ [deg.]\hhhu &
 $20.91^{+2.30}_{-2.57}$ & $[-0.2~10^{-2}] \pm 9.09$ & $3.76^{+6.04}_{-7.59}$ &
 $5.55^{+5.22}_{-6.81} $ & $7.45^{+4.50}_{-5.76}$ \\
\hline
$\gamma$ [deg.] \hhhu &
--$18.70 \pm 3.67$ & $[-.56~10^{-4}]\pm11.87$& $23.67^{+6.23}_{-4.87}$ &
$33.51^{+4.78}_{-3.67}$ & $44.95^{+3.32}_{-2.64}$ \\
\hline
\hline
$\theta_P$ [deg.] \hhhu &
--$8.75 \pm 1.01$ & --$11.33 \pm0 0.73$& --$10.07 \pm 0.79$
&--$9.70 \pm 0.72$& --$8.23 \pm 0.61$ \\
\hline
\hline
$\chi^2/dof$ \hhhu & 148.5/9 &  11.9/9 & 10.9/9 &10.9/9 & 10.9/9\\
\hline
\hline
\end{tabular}
 

\caption{
\label{two}
  Main fit  results fixing the
nonet symmetry violation parameter $x$ to various fixed values.      
The BKY parameter is fixed to standard SU(3) breaking $\ell_A=1.5$, the 
BGP parameter
is fixed to its no--breaking value $\ell_A=1.0$ and the process     
$K^{*+} \rightarrow K^+ \gamma$ is outside all fits.
Parameter values written boldface, means
 that they are not allowed to vary.
}
\newpage 
 
\begin{tabular}{|| c  | c  | c | c | c ||}
\hline
\hline
\hhhc  \hhhb  Process      & Nonet Sym. & Glue      & $K^{*\pm}$ &  PDG \\
\hhhc  \hhhb               & $+$ SU(3)  & $+$ SU(3) & Breaking   &  \\
\hline
\hline
$\rho \rightarrow \pi^0 \gamma$  $(\times 10^4)$ &\hhhb $5.16 \pm 0.03$&$5.16 \pm 0.03$ & $5.16 \pm 0.03$
& $6.8 \pm 1.7$\\
\hline
$\rho \rightarrow \pi^\pm \gamma $  $(\times 10^4)$ & $5.12 \pm 0.03$ & $5.12 \pm 0.03$ & $5.12 \pm 0.03$
& $4.5 \pm 0.5$ \hhhu \\
\hline
\hline
$\rho \rightarrow  \eta \gamma $  $(\times 10^4)$ \hhhu & $3.25 \pm 0.10$ & $3.28 \pm 0.10$ &$3.31 \pm 0.09$
& $2.4^{+0.8}_{-0.9}$\\
\hline
$\eta' \rightarrow \rho \gamma$ $(\times 10^2)$\hhhu & $33.1 \pm 2.0$ & $33.7 \pm 2.0$ & $33.0 \pm 1.8$
& $30.2 \pm 1.3$ \\
\hline
\hline
$K^{*\pm} \rightarrow K^\pm \gamma $  $(\times 10^4)$ \hhhu  & $5.66 \pm 0.03$ & $5.66 \pm 0.03$ &$9.80 \pm 0.93$
& $9.9 \pm 0.9$ \\
\hline
$K^{*0} \rightarrow K^0 \gamma $  $(\times 10^3)$ \hhhu  & $2.30 \pm 0.01$ & $2.30 \pm 0.01$ &$2.32 \pm 0.02$
& $2.3 \pm 0.2$\\
\hline
\hline
$\omega \rightarrow \pi^0 \gamma$ $(\times 10^2)$\hhhu  & $8.50 \pm 0.05$ & $8.50 \pm 0.05$ & $8.50 \pm 0.05$
& $8.5 \pm 0.5$ \\
\hline
$\omega \rightarrow \eta \gamma$  $(\times 10^4)$ \hhhu  & $8.0 \pm 0.2$ & $8.1 \pm 0.2$ & $8.12 \pm 0.19$ 
& $6.5 \pm 1.0$\\
\hline
$\eta' \rightarrow \omega \gamma$ $(\times 10^2)$\hhhu  & $2.8 \pm 0.2$ & $2.9 \pm 0.2$ & $2.8 \pm 0.2$
& $3.01 \pm 0.30$ \\
\hline
\hline
$\phi \rightarrow \pi^0 \gamma $  $(\times 10^3)$ \hhhu & $1.27 \pm 0.13$ & $1.28 \pm 0.12$ & $1.26 \pm 0.13$ 
& $1.31 \pm 0.13$\\
\hline
$\phi \rightarrow \eta \gamma$ $(\times 10^2)$\hhhu  & $1.25 \pm 0.04$ & $1.25 \pm 0.05$ & $1.22 \pm 0.04$
& $1.26 \pm 0.06$ \\
\hline
$\phi \rightarrow \eta' \gamma $  $(\times 10^4)$ \hhhu  & $0.61 \pm 0.027$ & $0.55 \pm 0.03$ & $0.63 \pm 0.02$
&$1.2^{+0.7}_{-0.5}$\\
\hline
\hline
$\eta \rightarrow \gamma \gamma$ $(\times 10^2)$\hhhu  & $40.5 \pm 1.7$ & $40.8 \pm 1.8$ & $41.5 \pm 1.4$
& $39.21 \pm 0.34$ \\
\hline
$\eta' \rightarrow \gamma \gamma$ $(\times 10^2)$\hhhu  & $2.1 \pm 0.1$ & $2.1 \pm 0.1$ & $2.1 \pm 0.1$
& $2.11 \pm 0.13$ \\
\hline
\hline
\end{tabular}

\caption{
\label{three}
Branching fractions from fits for radiative decays under various conditions of symmetry breakings.
Note that the rate for $K^{*\pm}$ is a prediction in the first two data columns,
while the corresponding data is included in the fit which  leads to the third data column.}

\newpage

\begin{tabular}{|| c  | c  | c | c | c ||}
\hline
\hline
\hhhc    Process     & Nonet Sym.  & Glue  & $K^{*\pm}$ & PDG \\
\hhhc                & $+$ SU(3)  & $+$ SU(3) & Breaking & \\
\hline
\hline
$\rho \rightarrow e^+ e^- $  $(\times 10^5)$ \hhhu & $4.6 \pm 0.1$ & $4.6 \pm 0.1$ & $4.6 \pm 0.1$
& $4.49 \pm 0.22$\\
\hline
\hline
$\omega \rightarrow e^+ e^- $  $(\times 10^5)$ \hhhu &  $7.0 \pm 0.2$ & $7.0 \pm 0.2$ & $7.0 \pm 0.2$
& $7.07 \pm 0.19$\\
\hline
\hline
$\phi \rightarrow e^+ e^- $  $(\times 10^4)$ \hhhu & $3.10 \pm 0.16$ & $3.11 \pm 0.16$ & $3.08 \pm 0.16$
& $2.99 \pm 0.08$\\
\hline
\hline
$\phi \rightarrow K^\pm K^\mp$ $(\times 10^2)$\hhhu &  $52.3 \pm 2.8$ & $52.3 \pm 2.7$ & $52.3 \pm 2.8$ 
& $49.1 \pm 0.8$ \\
\hline
\hline
$\phi \rightarrow K^0 \bar{K^0}$ $(\times 10^2)$\hhhu &  $34.0 \pm 1.7$ & $34.0 \pm 1.7$ & $34.1 \pm 1.7$ 
& $34.1 \pm 0.6$ \\
\hline
\hline
\end{tabular}

\caption{
\label{four}
Branching fractions from fits under various conditions of symmetry breakings. Note that
hadronic branching fractions for $\phi$ are predictions and that the corresponding
experimental value do not play any role in these predictions.
}

\end{table}
 

\appendix
\section{}

\subsection{The Standard Broken   $VVP$ Lagrangian}

\indent \indent Expressed in terms of $\ell_A$ and $\ell_W$ defined in the body of the text,
the SU(3) broken FKTUY  Lagrangian of relevance for us~: 

\begin{equation}
{\cal L}=- \frac{3 g^2}{4 \pi^2 f_{\pi}} 
\epsilon^{\mu \nu \alpha \beta} {\rm Tr}[\pa_\mu V_\nu X_W \pa_\alpha V_\beta X_A^{-1/2}P'X_A^{-1/2}]
=C\epsilon^{\mu \nu \alpha \beta} F_{\mu \nu \alpha \beta}
\label{lagwz0}
\end{equation}

\noindent can be developed in the following way~:

\bea\non
 F_{\mu\nu\alpha\beta}&=&
\pa_\mu\rho^0_\nu\left[\frac{1}{2\sqrt{\ell_A}}(
\pa_\alpha K^{*-}_\beta K^++\pa_\alpha K^{*+}_\beta K^--
\pa_\alpha \bar{K}^{*0}_\beta K^0-\pa_\alpha K^{*0}_\beta \bar{K}^0
)\right.\\
\non
&&\hspace{1cm}
\left.+\pa_\alpha\omega_\beta\pi^0+\pa_\alpha\rho^0_\beta
\left(\frac{\pi^8}{2\sqrt{3}}+\frac{x\eta_0}{\sqrt{6}}\right)\right]
\\
\non
&+&\pa_\mu\omega_\nu\left[
\frac{1}{2\sqrt{\ell_A}}(\pa_\alpha K^{*-}_\beta K^++\pa_\alpha K^{*+}_\beta K^-
+\pa_\alpha \bar{K}^{*0}_\beta K^0+\pa_\alpha K^{*0}_\beta \bar{K}^0
)\right.\\
\non
&&\hspace{1cm}
\left.+(\pa_\alpha\rho^+_\beta\pi^-+\pa_\alpha\rho^-_\beta\pi^+)
+\pa_\alpha\omega_\beta
\left(\frac{\pi^8}{2\sqrt{3}}+\frac{x\eta_0}{\sqrt{6}}\right)\right]
\\ \non
&+&\pa_\mu\phi_\nu\left[
\frac{\ell_W}{\sqrt{2 \ell_A}}(\pa_\alpha K^{*-}_\beta K^+
+\pa_\alpha K^{*+}_\beta K^-
+\pa_\alpha \bar{K}^{*0}_\beta K^0+\pa_\alpha K^{*0}_\beta \bar{K}^0
)\right.\\
\non
&&\hspace{1cm}
\left.
+\frac{\ell_W}{\ell_A}\pa_\alpha\phi_\beta
\left(-\frac{1}{\sqrt{3}}\pi^8
+\frac{1}{\sqrt{6}}x\eta_0\right)\right]
\\ \non
&+&\frac{1}{\sqrt{2 \ell_A}}\left[
\pa_\mu\rho^+_\nu(\pa_\alpha K^{*0}_\beta K^-+\pa_\alpha K^{*-}_\beta K^0)
+\pa_\mu\rho^-_\nu(
\pa_\alpha \bar{K}^{*0}_\beta K^++\pa_\alpha K^{*+}_\beta \bar{K}^0)
\right]
 \\
\non
&+&2\pa_\mu\rho^+_\nu\pa_\alpha\rho^-_\beta    
\left(\frac{\pi^8}{2\sqrt{3}}+\frac{x\eta_0}{\sqrt{6}}\right)
\\
\non
&+&\frac{\ell_W}{\sqrt{2}}
(\pa_\mu{K^{*-}_\nu}\pa_\alpha{K}^{*0}_\beta\pi^+
+\pa_\mu{K^{*+}_\nu}\pa_\alpha\bar{K}^{*0}_\beta\pi^-)
\\
\non
&+& \pa_\mu{K^{*-}_\nu}\pa_\alpha{K}^{*+}_\beta
\left[\frac{1}{2}\pi^0
-\frac{2-R}{2R\sqrt{3}}\pi^8
+\frac{1+R}{R\sqrt{6}}x\eta_0\right]
\\
\non
&+&\pa_\mu \bar{K}^{*0}_\nu\pa_\alpha K^{*0}_\beta
\left[-\frac{1}{2}\pi^0
-\frac{2-R}{2R\sqrt{3}} \pi^8
+\frac{1+R}{R\sqrt{6}}x\eta_0\right].\\
&&
\label{lagwz}
\eea

\noindent where $R=\ell_A \ell_W$ and $C=-3g^2/(8 \pi^2 f_{\pi})$. The expression
for the  functional $F$ in terms of the physical fields $\omega$, $\phi$,
$\eta$ and $\eta'$ can be trivially deduced.

\subsection{Matrix Elements with SU(3) and Nonet Broken Symmetries}

\indent \indent In terms of the angles ($\delta_V, ~\delta_A$) of physical
states with respect to ideal mixing, as defined in Section \ref{HLSmodel},
the coupling constants at vertices $VP\gamma$ which can be deduced from the 
Lagrangian in Eq. (\ref{lagwz}) are~:

\begin{equation}
\left \{
\begin{array}{lll}
G_{\rho^0 \pi^0 \gamma}=& &\displaystyle \frac{1}{3} G \\[0.3cm]
G_{\rho^{\pm} \pi^{\pm} \gamma}=& &\displaystyle \frac{1}{3} G \\[0.3cm]
G_{K^{*0} K^0 \gamma}=&- & \displaystyle G \frac{\sqrt{K}}{3} (1+\ell_W)  \\[0.3cm]
G_{K^{*\pm} K^{\pm} \gamma}=& &\displaystyle G \frac{\sqrt{K}}{3}(2-\ell_W)  \\[0.3cm]
G_{\rho^0 \eta \gamma}=& &\displaystyle \frac{1}{3} G
\left[\sqrt{2}(1-x)\cos{\delta_P}-(2x+1)\sin{\delta_P}\right]
\\[0.3cm]
G_{\rho^0 \eta' \gamma}=& &\displaystyle \frac{1}{3} G
\left[\sqrt{2}(1-x)\sin{\delta_P}+(2x+1)\cos{\delta_P}\right]\\[0.3cm]
G_{\omega \pi^0 \gamma}=& & \displaystyle G
\cos{\delta_V} \\[0.3cm]
G_{\phi \pi^0 \gamma}=&- & \displaystyle G \sin{\delta_V}
\end{array}
\right.
\label{cc1}
\end{equation}

\noindent and~:

\begin{equation}
\left \{
\begin{array}{lll}
G_{\omega \eta \gamma}=& & \displaystyle \frac{1}{9} G \left [
-(2x+1) \cos{\delta_V}\sin{\delta_P} 
-2 Z(2+x)\sin{\delta_V}\cos{\delta_P} \right .\\[0.3cm]
~~~& & \displaystyle \left .
+2\sqrt{2}Z(1-x)\sin{\delta_V}\sin{\delta_P}
+\sqrt{2}(1-x)\cos{\delta_V}\cos{\delta_P}\right ]\\[0.3cm]

G_{\omega \eta' \gamma}=&&\displaystyle \frac{1}{9} G \left [
(2x+1)\cos{\delta_V}\cos{\delta_P} 
-2Z(2+x)\sin{\delta_V}\sin{\delta_P} \right .\\[0.3cm]
~~~&& \displaystyle \left .
-2\sqrt{2}Z(1-x)\sin{\delta_V}\cos{\delta_P}
+\sqrt{2}(1-x)\cos{\delta_V}\sin{\delta_P} \right ]\\[0.3cm]

G_{\phi \eta \gamma}=&&\displaystyle \frac{1}{9} G \left [
(2x+1)\sin{\delta_V}\sin{\delta_P} 
-2Z(2+x)\cos{\delta_V}\cos{\delta_P}\right . \\[0.3cm]
~~~&& \displaystyle \left .
+2\sqrt{2}Z(1-x)\cos{\delta_V}\sin{\delta_P}
-\sqrt{2}(1-x)\sin{\delta_V}\cos{\delta_P} \right ]\\[0.3cm]

G_{\phi \eta' \gamma}=&-&\displaystyle \frac{1}{9}G \left [
(2x+1)\sin{\delta_V}\cos{\delta_P} 
+2Z(2+x)\cos{\delta_V}\sin{\delta_P}  \right .\\[0.3cm]
~~~&& \displaystyle \left .
+2\sqrt{2}Z(1-x)\cos{\delta_V}\cos{\delta_P}
+\sqrt{2}(1-x)\sin{\delta_V}\sin{\delta_P} \right ]
\end{array}
\right.
\label{cc2}
\end{equation}

\noindent where the breaking parameters come through
 the combinations $K=1/\ell_A$ and $Z=\ell_W/\ell_A$.
The dependence upon the nonet symmetry breaking parameter $x$ 
is explicit. The basic parameter $G$ yields the following expression~:

\begin{equation}
 G=\frac{Ce}{2g}=-\frac{3eg}{8 \pi^2 f_{\pi}}
\label{cc3}
\end{equation}

\noindent which depends on the vector meson universal coupling
$g$ and the pion decay constant $f_{\pi}$. We will fit the
absolute value of $G$. 
 
 Correspondingly the matrix elements for the decays
$\pi^0/\eta/\eta' \rightarrow \gamma \gamma$ are~:
 
\begin{equation}
\left \{
\begin{array}{lll}
G_{\eta \gamma \gamma} = && -\displaystyle \frac{\alpha_{em}}{\pi \sqrt{3} f_{\pi}}
\left [ \frac{5-2Z}{3}\cos{\theta_P}-\sqrt{2} \frac{5+Z}{3}x \sin{\theta_P} \right ]\\[0.3cm] 
G_{\eta' \gamma \gamma} = && -\displaystyle \frac{\alpha_{em}}{\pi \sqrt{3} f_{\pi}}
\left [ \frac{5-2Z}{3}\sin{\theta_P} + \sqrt{2} \frac{5+Z}{3}x \cos{\theta_P} \right ]\\[0.3cm] 
G_{\pi^0 \gamma \gamma} = && -\displaystyle  \frac{\alpha_{em}}{\pi  f_{\pi}}
\end{array}
\right.
\label{cc4}
\end{equation}

Finally the $V-\gamma$ couplings which enter Eq. (\ref{brk5}) 
are constant within this framework and are~:

\begin{equation}
\left \{
\begin{array}{lll}
f_{\rho \gamma} = && \displaystyle a f_{\pi}^2 g \\[0.3cm] 
f_{\omega \gamma} = && \displaystyle \frac{f_{\rho \gamma}}{3}
\left [ \cos{\delta_V}+\ell_V \sqrt{2}\sin{\delta_V}\right ]\\[0.3cm] 
f_{\phi\gamma} = &-& \displaystyle  \frac{f_{\rho \gamma}}{3}
\left [ \sin{\delta_V}-\ell_V \sqrt{2}\cos{\delta_V}\right ]
\end{array}
\right.
\label{cc5}
\end{equation}

We have defined above $K$ and $Z$ in terms of the breaking parameters
of Bando Kugo Yamawaki (BKY) \cite{BKY}, and of that of
Bramon, Grau and Pancheri (BGP) \cite{BGP}~:

\begin{equation}
\begin{array}{lll}
\ell_A=1+c_A~~~, &~~~\ell_W=1+c_W &~~~\ell_V=(1+c_V~)^2~~.
\end{array}
\label{brkrel}
\end{equation}

\subsection{Matrix Elements with All Breakings and Coupling to Glue}

By inverting Eq. (\ref{model2}), we get
expressions for $\pi_8$ and $\eta_0$ in terms of
the three mixing angles ($\theta_{P}, ~\beta, ~\gamma$)
and of the $\eta$ and $\eta'$ fields\footnote{We ignore
the spurious $\eta''$ as long as we don't know what
particle it could correspond to.}. {From} these expressions
and the phenomenological Lagrangian of Eq. (\ref{lagwz}),
we can recompute the expressions corresponding to Eqs.
(\ref{cc1}), (\ref{cc2}) and (\ref{cc4}). 

Of course, all $G_{VP\gamma}$'s, where $P$ is neither one of
$\eta$ and $\eta'$, are unchanged and coincide with the corresponding 
expressions given in Eqs. (\ref{cc1}). The other expressions
become~:

\begin{equation}
\left \{
\begin{array}{lll}
G_{\rho^0 \eta \gamma}=& &\displaystyle \frac{\cos{\beta}}{3} G
\left \{\sqrt{2}(1-x)\cos{\delta_P}-(2x+1)\sin{\delta_P}\right\}
\\[0.4cm]

G_{\rho^0 \eta' \gamma}=& &\displaystyle \frac{\cos{\gamma}}{3} G
\left\{
[\sqrt{2}(1-x)+(2x+1)\sin{\beta}\tan{\gamma}]\tan{\delta_P} \right.\\[0.4cm]
 ~~~&&\left . +[(2x+1)-\sqrt{2}(1-x)\sin{\beta}\tan{\gamma}]\cos{\delta_P}\right\}\\
\end{array}
\right.
\label{cc6}
\end{equation}

\begin{equation}
\left \{
\begin{array}{lll}
G_{\omega \eta \gamma}=& & \displaystyle \frac{\cos{\beta}}{9} G \left \{
-(2x+1) \cos{\delta_V}\sin{\delta_P} 
-2 Z(2+x)\sin{\delta_V}\cos{\delta_P} \right .\\[0.3cm]
~~~& & \displaystyle \left .
+2\sqrt{2}Z(1-x)\sin{\delta_V}\sin{\delta_P}
+\sqrt{2}(1-x)\cos{\delta_V}\cos{\delta_P}\right \}\\[0.3cm]

G_{\omega \eta' \gamma}=&&\displaystyle \frac{\cos{\gamma}}{9} G \left \{
[(2x+1)-\sqrt{2}(1-x)\sin{\beta}\tan{\gamma}]
\cos{\delta_V}\cos{\delta_P} \right.\\[0.3cm]
~~&&-2Z[(2+x)+\sqrt{2}(1-x)\sin{\beta}\tan{\gamma}]
\sin{\delta_V}\sin{\delta_P} \\[0.3cm]
~~~&& \displaystyle 
-2Z[\sqrt{2}(1-x)-(2+x)\sin{\beta}\tan{\gamma})]\sin{\delta_V}\cos{\delta_P}\\[0.3cm]
~&&\left .
+[\sqrt{2}(1-x)+(2x+1)\sin{\beta}\tan{\gamma}]\cos{\delta_V}\sin{\delta_P} \right \}\\[0.3cm]

G_{\phi \eta \gamma}=&&\displaystyle \frac{\cos{\beta}}{9} G \left \{
(2x+1)\sin{\delta_V}\sin{\delta_P} 
-2Z(2+x)\cos{\delta_V}\cos{\delta_P}\right . \\[0.3cm]
~~~&& \displaystyle \left .
+2\sqrt{2}Z(1-x)\cos{\delta_V}\sin{\delta_P}
-\sqrt{2}(1-x)\sin{\delta_V}\cos{\delta_P} \right \}\\[0.3cm]

G_{\phi \eta' \gamma}=&-&\displaystyle \frac{\cos{\gamma}}{9}G \left \{
[(2x+1)-\sqrt{2}(1-x)\sin{\beta}\tan{\gamma}]\sin{\delta_V}\cos{\delta_P} \right.\\[0.3cm]
~&&
+2Z[(2+x)+\sqrt{2}(1-x)\sin{\beta}\tan{\gamma}]\cos{\delta_V}\sin{\delta_P}  \\[0.3cm]
~~~&& \displaystyle 
+2Z[\sqrt{2}(1-x)-(2+x)\sin{\beta}\tan{\gamma}]\cos{\delta_V}\cos{\delta_P}  \\[0.3cm]
~~~&& \displaystyle \left .
+[\sqrt{2}(1-x)+(2x+1)\sin{\beta}\tan{\gamma}]\sin{\delta_V}\sin{\delta_P} \right \}
\end{array}
\right.
\label{cc7}
\end{equation}

 Correspondingly the matrix element for the decay $\pi^0 \rightarrow \gamma \gamma$ 
 is not affected, while the amplitudes for   $\eta/\eta' \rightarrow \gamma \gamma$
 become~:
\begin{equation}
\left \{
\begin{array}{lll}
G_{\eta \gamma \gamma} = && -\displaystyle \frac{\alpha_{em} \cos{\beta}}{\pi \sqrt{3} f_{\pi}}
\left \{ \frac{5-2Z}{3}\cos{\theta_P}-\sqrt{2} \frac{5+Z}{3}x \sin{\theta_P} \right \}\\[0.4cm]
 
G_{\eta' \gamma \gamma} = && -\displaystyle 
\frac{\alpha_{em}\cos{\gamma}}{\pi \sqrt{3} f_{\pi}}
\left \{ [\frac{5-2Z}{3}+\sqrt{2} \frac{5+Z}{3}x\sin{\beta}\tan{\gamma} ] 
\sin{\theta_P} \right.\\[0.4cm]
~&&\displaystyle \left.
+[\sqrt{2} \frac{5+Z}{3}x -\frac{5-2Z}{3}\sin{\beta}\tan{\gamma}]\cos{\theta_P} \right \} 
\end{array}
\right.
\label{cc8}
\end{equation}

\subsection{The $K^*$ Broken   $VVP$ Lagrangian}

\indent \indent In order to describe fully the $K^*$ sector
of light meson radiative decays, we propose to use
the Lagrangian given in Eq. (\ref{lagwz0}),
supplemented with the replacement $V \rightarrow X_T V X_T$,
with

\begin{equation}
X_T=\rm{diag}(1~,1~,1+c_T)
\label{de1}
\end{equation}
\noindent where the breaking parameter, expected to be small,
has been denoted $c_T$. In the following we use the definition

\begin{equation}
\ell_T=(1+c_T)^2
\label{de2}
\end{equation}
\noindent together with the breaking parameters already defined
by Eqs. (\ref{brkrel}) and the nonet symmetry breaking factor $x$.

The expanded expression for the FKTUY Lagrangian broken in this way is
still given by Eq.~(\ref{lagwz0}) with

\bea\non
 F_{\mu\nu\alpha\beta}&=&
\pa_\mu\rho^0_\nu\left[\frac{1}{2}\sqrt{\frac{\ell_T}{\ell_A}}(
\pa_\alpha K^{*-}_\beta K^++\pa_\alpha K^{*+}_\beta K^--
\pa_\alpha \bar{K}^{*0}_\beta K^0-\pa_\alpha K^{*0}_\beta \bar{K}^0
)\right.\\
\non
&&\hspace{1cm}
\left.+\pa_\alpha\omega_\beta\pi^0+\pa_\alpha\rho^0_\beta
\left(\frac{\pi^8}{2\sqrt{3}}+\frac{x\eta_0}{\sqrt{6}}\right)\right]
\\
\non
&+&\pa_\mu\omega_\nu\left[
\frac{1}{2}\sqrt{\frac{\ell_T}{\ell_A}}
(\pa_\alpha K^{*-}_\beta K^++\pa_\alpha K^{*+}_\beta K^-
+\pa_\alpha \bar{K}^{*0}_\beta K^0+\pa_\alpha K^{*0}_\beta \bar{K}^0
)\right.\\
\non
&&\hspace{1cm}
\left.+(\pa_\alpha\rho^+_\beta\pi^-+\pa_\alpha\rho^-_\beta\pi^+)
+\pa_\alpha\omega_\beta
\left(\frac{\pi^8}{2\sqrt{3}}+\frac{x\eta_0}{\sqrt{6}}\right)\right]
\\ \non
&+&\pa_\mu\phi_\nu\left[
\ell_W\sqrt{\frac{\ell_T^3}{2 \ell_A}}
(\pa_\alpha K^{*-}_\beta K^++\pa_\alpha K^{*+}_\beta K^-
+\pa_\alpha \bar{K}^{*0}_\beta K^0+\pa_\alpha K^{*0}_\beta \bar{K}^0
)\right.\\
\non
&&\hspace{1cm}
\left.
+\frac{\ell_W \ell_T^2}{\ell_A}\pa_\alpha\phi_\beta
\left(-\frac{1}{\sqrt{3}}\pi^8
+\frac{1}{\sqrt{6}}x\eta_0\right)\right]
\\ \non
&+&\sqrt{\frac{\ell_T}{2 \ell_A}}\left[
\pa_\mu\rho^+_\nu(\pa_\alpha K^{*0}_\beta K^-+\pa_\alpha K^{*-}_\beta K^0)
+\pa_\mu\rho^-_\nu(
\pa_\alpha \bar{K}^{*0}_\beta K^++\pa_\alpha K^{*+}_\beta \bar{K}^0)
\right]
 \\
\non
&+&2\pa_\mu\rho^+_\nu\pa_\alpha\rho^-_\beta    
\left(\frac{\pi^8}{2\sqrt{3}}+\frac{x\eta_0}{\sqrt{6}}\right)
\\
\non
&+&\frac{\ell_W \ell_T}{\sqrt{2}}
(\pa_\mu{K^{*-}_\nu}\pa_\alpha{K}^{*0}_\beta\pi^+
+\pa_\mu{K^{*+}_\nu}\pa_\alpha\bar{K}^{*0}_\beta\pi^-)
\\
\non
&+& \ell_T\pa_\mu{K^{*-}_\nu}\pa_\alpha{K}^{*+}_\beta
\left[\frac{1}{2}\pi^0
-\frac{2-R}{2R\sqrt{3}}\pi^8
+\frac{1+R}{R\sqrt{6}}x\eta_0\right]
\\
\non
&+&\ell_T\pa_\mu \bar{K}^{*0}_\nu\pa_\alpha K^{*0}_\beta
\left[-\frac{1}{2}\pi^0
-\frac{2-R}{2R\sqrt{3}} \pi^8
+\frac{1+R}{R\sqrt{6}}x\eta_0\right].\\
&&
\label{lagwzb}
\eea

\noindent where $R=\ell_A \ell_W$, as in Eq. (\ref{lagwz}).

\end{document}